\documentclass[a4paper,11pt]{article}

\pdfoutput=1 

\usepackage{jcappub} 
\usepackage{amsmath}
\usepackage{amsfonts}
\usepackage{amssymb}
\usepackage{graphicx}
\def\figurename{Fig.}

\title{\boldmath New slow-roll approximations for inflation in Einstein-Gauss-Bonnet gravity}

\author[a]{Ekaterina~O.~Pozdeeva,}
\author[b]{Maria~A.~Skugoreva,}
\author[b,c]{Alexey~V.~Toporensky,}
\author[a]{Sergey~Yu.~Vernov}

\affiliation[a]{Skobeltsyn Institute of Nuclear Physics, Lomonosov Moscow State University,\\
Leninskiye Gory~1, Moscow 119991, Russia}
\affiliation[b]{Kazan Federal University, Kremlevskaya Street~18, Kazan 420008, Russia}
\affiliation[c]{Sternberg Astronomical Institute, Lomonosov Moscow State University, Universitetsky Prospect~13, Moscow 119991, Russia}

\emailAdd{pozdeeva@www-hep.sinp.msu.ru}
\emailAdd{masha-sk@mail.ru}
\emailAdd{atopor@rambler.ru}
\emailAdd{svernov@theory.sinp.msu.ru}


\arxivnumber{2403.06147}

\abstract{We propose new slow-roll approximations for inflationary models with the Gauss-Bonnet term. We find more accurate expressions of the standard slow-roll parameters as functions of the scalar field. To check the accuracy of approximations considered we construct inflationary models with quadratic and quartic monomial potentials and the Gauss-Bonnet term. Numerical analysis of these models indicates that the proposed inflationary scenarios do not contradict to the observation data. New slow-roll approximations show that the constructed inflationary models are in agreement with the observation data, whereas one does not get allowed observational parameters at the same values of parameters of the constructed models in the standard slow-roll approximation.
}

\begin{document}
\maketitle
\flushbottom

\section{Introduction}
~~~~Recent progress in observations of relic cosmological perturbations already makes possible to severely constraint inflationary theories~\cite{Planck:2018jri,BICEP:2021xfz,Galloni:2022mok}, ruling out inflationary models with a minimally coupled scalar field and monomial potentials. Amplitude of scalar perturbations $A_s$ and their spectral index $n_s$ are known with a good accuracy, as for the tensor-to-scalar ratio $r$, only upper bound is known, being however more and more strict with the progress in observation technics. The main inflationary parameters are constrained by the combined analysis of Planck, BICEP/Keck and other observations as follows~\cite{Galloni:2022mok}:
\begin{equation}
\label{Inflparamobserv}
A_s=(2.10\pm 0.03)\times 10^{-9},\qquad n_s=0.9654\pm 0.0040 \qquad {\rm and} \qquad  r < 0.028 \,.
\end{equation}

    To calculate these inflationary parameters one usually uses the slow-roll approximation. The slow-roll parameters are defined as functions of the Hubble parameter and the scalar field. For the General Relativity (GR) models with minimally coupling scalar fields, the slow-roll approximation is equivalent to the assumption that all slow-roll parameters are small in comparison with unity. In this approximation, the Hubble parameter and slow-roll parameters are functions of the scalar field only~(see, for example,~Ref.~\cite{Liddle:1994dx}). However, already for a nonminimally coupled scalar field, the correct description of slow-roll approximation appears to be more subtle~\cite{Kaiser:1994vs,Akin:2020mcr, Jarv:2021qpp,Diaz:2023tma}.

    The slow-roll approximation essentially simplifies the search for model parameters suitable for description of the inflation. If we know $n_s(\phi)$, $r(\phi)$, and $A_s(\phi)$, then we can calculate the value of $\phi$ and parameters of the model considered getting inflationary parameters that do not contradict to the observation data~\eqref{Inflparamobserv}. Also, the slow-roll approximation is useful to get an approximate value of $\phi$ that corresponds to the end of inflation.

    In this paper, we consider slow-roll approximations for models with the Gauss-Bonnet term, described by the following action:
\begin{equation}
\label{action1}
S=\int d^4x\sqrt{-g}\left[U_0R-\frac{1}{2}g^{\mu\nu}\partial_\mu\phi\partial_\nu\phi-V(\phi)-\frac{1}{2}\xi(\phi){\cal G}\right],
\end{equation}
where $U_0>0$ is a constant, the functions $V(\phi)$ and $\xi(\phi)$ are differentiable ones, $R$ is the Ricci scalar and
\begin{equation*}
\mathcal{G}=R_{\mu\nu\rho\sigma}R^{\mu\nu\rho\sigma}-4R_{\mu\nu}R^{\mu\nu}+R^2
\end{equation*}
is the Gauss-Bonnet term. Note that the Gauss-Bonnet term coupled to a scalar field arises naturally in string inspired cosmological models~\cite{Antoniadis:1993jc,Torii:1996yi,Kawai:1998ab,Kawai:1999pw,Cartier:2001is,Hwang:2005hb,Sami:2005zc,Tsujikawa:2006ph}. Cosmological perturbations in models with the Gauss-Bonnet term have been explored in Refs.~\cite{Kawai:1998ab,Kawai:1999pw,Cartier:2001is,Hwang:2005hb}.

    Inflationary models with the Gauss-Bonnet term are popular~\cite{Guo:2009uk,Guo:2010jr,Jiang:2013gza,Koh:2014bka,Hikmawan:2015rze,vandeBruck:2015gjd,vandeBruck:2016xvt,Koh:2016abf,Mathew:2016anx,Chakraborty:2018scm,Yi:2018gse,Odintsov:2018zhw,Koh:2018qcy,Kleidis:2019ywv,Odintsov:2019clh,Rashidi:2020wwg,Pozdeeva:2020shl,Pozdeeva:2020apf,Pozdeeva:2021iwc,Oikonomou:2021kql,Pozdeeva:2021nmz,Younesizadeh:2021heg,Kawai:2021bye,Kawai:2021edk,Odintsov:2022zrj,Oikonomou:2022ksx,Kawaguchi:2022nku,Gangopadhyay:2022vgh,Odintsov:2023aaw,Nojiri:2023mvi,Odintsov:2023lbb,Mudrunka:2023wxy,Oikonomou:2024jqv}.
 Many of them include the function $\xi$ inversely proportional to the scalar field potential: $\xi(\phi)=\xi_0/V(\phi)$, where $\xi_0$ is a constant~\cite{Guo:2009uk,Guo:2010jr,Jiang:2013gza,Koh:2016abf,Yi:2018gse,Odintsov:2018zhw,Kleidis:2019ywv,Rashidi:2020wwg,Pozdeeva:2020shl}.

    The slow-roll approximation of such models includes two additional slow-roll parameters. The standard slow-roll approximation proposed in Ref.~\cite{Guo:2010jr} uses the condition that all inflationary parameters are negligibly small during inflation.

  In the case of a model with a monomial potential and the function $\xi(\phi)$ inversely proportional to the potential, this function rapidly raises at the end of inflation, when $\phi$ tends to zero, and there are problems with the exit from inflation. It has been shown in Ref.~\cite{vandeBruck:2015gjd} by numerical calculations that the model with the fourth degree monomial potential $V=V_0\phi^4$ and $\xi=\xi_0/V$, where $V_0$ and $\xi_0$ are some positive constants, has no exit from inflation, whereas the standard slow-roll approximation shows that this exit does exist, hence, this approximation is not accurate at the end of inflation. So, it is important to improve the slow-roll approximation and compare approximate results with results of numerical calculations without any approximation.

    In this paper, we propose and compare two new  slow-roll approximations. For the standard slow-roll parameters proposed in Ref.~\cite{Guo:2010jr}, we obtain new expressions of these parameters as functions of $\phi$. To compare new approximations with the standard one and with numerical calculations without any approximation we consider models with quadratic and quartic potentials,  $V=V_0\phi^n$, where $V_0>0$ and $n=2,\,4$, and the function $\xi(\phi)=\xi_0/(V(\phi)+\Lambda)$, with  constants $\xi_0>0$ and $\Lambda>0$. So, $\xi(\phi)$ has no singularity, and the above-mentioned problem  of the exit from inflation disappears (see below).

    The paper is organized as follows. In Section~2, we remind the evolution equations and introduce the slow-roll parameters. In Section~3, we remind the standard slow-roll approximation. In Section~4, we propose new slow-roll approximations. Inflationary models with monomial potentials $V$ are investigated in Section~5. Our results are summarized in Section~6. Explicit expressions for the slow-roll parameters as functions of $\phi$ are presented in Appendix~A.

\section{Evolution equations for Einstein-Gauss-Bonnet gravity and slow-roll parameters}
~~~~In the spatially flat Friedmann-Lema\^{i}tre-Robertson-Walker metric with
\begin{equation*}
ds^2={}-dt^2+a^2(t)(dx^2+dy^2+dz^2),
\end{equation*}
one obtains the following system of evolution equations~\cite{vandeBruck:2015gjd,Pozdeeva:2020apf}:
\begin{eqnarray}
12H^2\left(U_0-2\xi_{,\phi}\psi H\right)&=&\psi^2+2V,
\label{Equ00} \\
4\dot H\left(U_0-2\xi_{,\phi}\psi H\right)&=&{}-\psi^2+4H^2\left(\xi_{,\phi\phi}\psi^2+\xi_{,\phi}\dot\psi-H\xi_{,\phi}\psi\right),
\label{Equ11} \\
\dot{\psi}+3H\psi&=&{} -V_{,\phi} -12H^2\xi_{,\phi}\left(\dot{H}+H^2\right),\label{Equphi}
\end{eqnarray}
where $H=\dot{a}/a$ is the Hubble parameter, $a(t)$ is the scale factor, $\psi=\dot{\phi}$, dots denote the derivatives with respect to the cosmic time $t$ and $A_{,\phi} \equiv \frac{dA}{d\phi}$ for any function $A(\phi)$.

It is easy to show that Eq.~(\ref{Equ11}) is a consequence of Eqs.~(\ref{Equ00}) and (\ref{Equphi}).

If $\xi_{,\phi}\neq 0$, then equations (\ref{Equ11}) and (\ref{Equphi}) do not form a dynamic system. It is suitable to use the following combinations of Eqs.~(\ref{Equ11}) and (\ref{Equphi}) that form a dynamical system~\cite{Pozdeeva:2019agu}:
\begin{equation}
\label{DynSYS}
\begin{split}
\dot\phi=&\psi,\\
\dot\psi=&\frac{1}{B-2\xi_{,\phi}H\psi}\left\{
3\left[3-4\,\xi_{,\phi\phi} H^2 \right]\xi_{,\phi}H^2\psi^2+\left[3B+2\,\xi_{,\phi}V_{,\phi}-6U_0\right]H\psi-\frac{V^2}{U_0}X\right\},\\
\dot H=&\frac{1}{4\left(B-2\,\xi_{,\phi}H\psi\right)}\left\{\left(4\,\xi_{,\phi\phi}H^2-1\right)\psi^2-16\xi_{,\phi}H^3\psi-4\frac{V^2}{U_0^2}\,\xi_{,\phi} H^2 X\right\},
\end{split}
\end{equation}
where
\begin{equation*}
B=12\xi_{,\phi}^2H^4+U_0,\qquad
X=\frac{U_0^2}{V^2}\left(12\xi_{,\phi} H^4+V_{,\phi} \right).
\end{equation*}

The third equation of system (\ref{DynSYS}) is a consequence of Eqs.~(\ref{Equ00}) and (\ref{Equphi}).

As usually for inflationary model construction, the e-folding number $N=\ln(a/a_{e})$, where $a_{e}$ is a constant, is considered as a measure of time during inflation.

 Using the relation $\frac{d}{dt}=H\,\frac{d}{dN}$ and introducing $\chi=\psi/H$, one can write system~(\ref{DynSYS}) as follows:
\begin{equation}
\label{DynSYSN}
\begin{split}
\frac{d\phi}{dN}=&\,\chi,\\
\frac{d\chi}{dN}=&\,\frac{1}{H^2\left(B-2\xi_{,\phi}H^2\chi\right)}\left\{
3\left[3-4\xi_{,\phi\phi} H^2\right]\xi_{,\phi}H^4\chi^2+ \left[3B+2\xi_{,\phi}V_{,\phi}-6U_0\right]H^2\chi-\frac{V^2}{U_0}X\right\}\\
&{}-\frac{\chi}{2H^2}\frac{dH^2}{dN},\\
\frac{dH^2}{dN}=&\,\frac{H^2}{2\left(B-2\xi_{,\phi}H^2\chi\right)}\left\{\left(4\xi_{,\phi\phi}H^2-1\right)\chi^2-16\xi_{,\phi}H^2\chi-4\frac{V^2}{U_0^2}\xi_{,\phi}X\right\}.
\end{split}
\end{equation}

   Equation (\ref{Equ00}) can be presented in the following form
\begin{equation}
\label{Equ00N}
24\chi\xi_{,\phi} H^4+\left(\chi^2- 12U_0\right)H^2+2V=0.
\end{equation}
and has solutions:
\begin{equation}
\label{Q0}
H^2_\pm=\frac{12U_0-\chi^2\pm\sqrt{\left(12U_0-\chi^2\right)^2-192V\chi\xi_{,\phi}}}{48\,\chi\,\xi_{,\phi}},
\end{equation}
if $\chi\,\xi_{,\phi}\neq 0$. In the opposite case, the solution is
\begin{equation}
H^2_0=\frac{2V}{12U_0-\chi^2}\,.
\end{equation}
Equation (\ref{Equ00N}) restricts the set of initial conditions of system~(\ref{DynSYSN}).

Following Refs.~\cite{Guo:2010jr,vandeBruck:2015gjd,Pozdeeva:2020apf,Odintsov:2023lbb}, we consider the slow-roll parameters:
\begin{equation}
\label{epsilon}
\varepsilon_1 ={}-\frac{\dot{H}}{H^2}={}-\frac{1}{2}\frac{d\ln(H^2)}{dN},\qquad \varepsilon_{i+1}= \frac{d\ln|\varepsilon_i|}{dN},\quad i\geqslant 1,
\end{equation}
\begin{equation}
\label{delta}
\delta_1= \frac{2}{U_0}\xi_{,\phi}H\psi=\frac{2}{U_0}\xi_{,\phi}H^2\chi,\qquad \delta_{i+1}=\frac{d\ln|\delta_i|}{dN}, \quad i\geqslant 1.
\end{equation}

    It is easy to see that
\begin{equation}
\label{delta2}
\delta_2=\frac{\dot{\psi}}{H\psi}+\frac{\xi_{,\phi\phi}\psi}{H\xi_{,\phi}}-\varepsilon_1.
\end{equation}

    Using system~(\ref{DynSYSN}), we obtain that the parameter $\varepsilon_1(N)$ satisfies the following equation:
\begin{equation}
\label{equepsilon1a}
\varepsilon_1={}-\frac{1}{4\left(B-2\xi_{,\phi}H^2\chi\right)}\left\{\left(4\xi_{,\phi\phi}H^2-1\right)\chi^2-16\xi_{,\phi}H^2\chi-4\frac{V^2}{U_0^2}\xi_{,\phi}X\right\}\,.
\end{equation}

    Dividing Eq.~(\ref{Equ11}) on Eq.~(\ref{Equ00}), we get another form of $\varepsilon_1$:
\begin{equation}
\label{equepsilon1b}
\varepsilon_1=\frac{3}{\psi^2+2V}\left[\psi^2-4H^2\left(\xi_{,\phi\phi}\psi^2+\xi_{,\phi}\dot\psi-H\xi_{,\phi}\psi\right)\right].
\end{equation}

    Note that we do not use any approximation, so, these expressions for $\varepsilon_1$ are exact.

    It is useful, to rewrite evolution equations in terms of the slow-roll parameters. Equations~(\ref{Equ00}) and (\ref{Equ11}) are equivalent to
\begin{equation}
\label{Equ00delta1}
12U_0H^2\left(1-\delta_1\right)=\psi^2+2V=H^2\chi^2+2V\,,
\end{equation}
\begin{equation}
\label{Equ11delta1}
4U_0\dot{H}\left(1-\delta_1\right)={}-\psi^2+2U_0H^2\delta_1\left(\delta_2+\varepsilon_1-1\right).
\end{equation}

    From Eq.~(\ref{Equ11delta1}), it follows
\begin{equation}
\label{chiviaslr}
\chi^2=2U_0\left[2\varepsilon_1-\delta_1+\delta_1\left(\delta_2-\varepsilon_1\right)\right]\,.
\end{equation}

    The spectral index $n_s$ and the tensor-to-scalar ratio $r$ are connected with the slow-roll parameters as follows~\cite{Guo:2010jr},
\begin{equation}
\label{ns_slr}
n_s=1-2\varepsilon_1-\frac{2\varepsilon_1\varepsilon_2-\delta_1\delta_2}{2\varepsilon_1-\delta_1}=1-2\varepsilon_1-\frac{d\ln(r)}{dN}=1+\frac{d}{dN}\ln\left(\frac{H^2}{U_0r}\right),
\end{equation}
\begin{equation}
\label{r_slr}
r=8|2\varepsilon_1-\delta_1|.
\end{equation}

    From Eq.~(\ref{chiviaslr}), it follows that
\begin{equation*}
r=\left|\frac{U_0}{\xi_{,\phi}^2H^4}\delta_1^2-8\delta_1\delta_2+8\delta_1\varepsilon_1\right|\,.
\end{equation*}

 The scalar perturbation amplitude is given as
\begin{equation}
\label{As_slr}
A_s=\frac{H^2}{\pi^2 U_0\, r}\,.
\end{equation}

Proposed in Refs.~\cite{Hwang:2005hb,Guo:2010jr,vandeBruck:2015gjd}, the general expressions for the scalar perturbation propagation speed

\begin{equation}
\label{cR}
c_{R}^2=1+\frac{8H^2{\dot\xi}^2\left[2\dot H\left(U_0-2H{\dot\xi}\right)+H^2\left(\ddot{\xi}- H{\dot\xi}\right)\right]}{\left(U_0-2H{\dot\xi}\right)
\left[{\dot\phi}^2\left(U_0-2H{\dot\xi}\right)+12 H^4{\dot\xi}^2\right]}
\end{equation}
and the gravitational wave propagation speed

\begin{equation}
\label{cT}
    c_{T}^2 = 1-\frac{2\left(\ddot{\xi}-H{\dot\xi}\right)}{U_0-2H{\dot\xi}}.
\end{equation}
can be expressed in terms of the slow-roll parameters as follows~\cite{Guo:2010jr}:
\begin{equation}
\label{cRepsdel}
c_{R}^2 = 1 - \frac{J^2}{K} \left(2\varepsilon_1+\frac12 \delta_1\left(1-5\varepsilon_1-\delta_2\right)\right),
\end{equation}
\begin{equation}
\label{cTepsdel}
c_{T}^2 = 1 + J\left(1-\varepsilon_1-\delta_2\right),
\end{equation}
where
\begin{equation}
J=\frac{\delta_1}{1-\delta_1},\qquad K=2\varepsilon_1-\delta_1\left(1+\varepsilon_1-\delta_2\right)+\frac32 J \delta_1\,.
\end{equation}

\section{The standard slow-roll approximation and the effective potential}
\label{slow-roll approximation}
~~~~There are a few ways to get the slow-roll approximate equations.  The standard approximate equations have been proposed in Ref.~\cite{Guo:2010jr}. This way assumes that all inflationary parameters are negligibly small and can be removed from equations. In this slow-roll approximation, the leading order equations have the following form:
\begin{eqnarray}
H^2&\simeq&\frac{V}{6U_0}\,,
\label{Equ0lo}\\
\dot H &\simeq &{}-\frac{\psi^2}{4U_0}-\frac{\xi_{,\phi}H^3\psi}{U_0}\,, \label{EquHlo}\\
\psi& \simeq &{}-\frac{V_{,\phi}+12\xi_{,\phi}H^4}{3H}.
\label{Equphilo}
\end{eqnarray}

New form of the slow-roll equations which uses the effective potential was developed in Ref.~\cite{Pozdeeva:2020apf}. The effective potential
\begin{equation}
\label{Veff}
V_{eff}(\phi)={}-\frac{U_0^2}{V(\phi)}+\frac{1}{3}\xi(\phi)
\end{equation}
has been proposed~\cite{Pozdeeva:2019agu} (see also,~\cite{Pozdeeva:2020apf,Vernov:2021hxo}) to analyze the stability of de Sitter solutions in model (\ref{action1}). The effective potential~\eqref{Veff} is not defined in the case of $V(\phi)\equiv 0$, but inflationary scenarios are always unstable in this case~\cite{Hikmawan:2015rze}. In this paper, we consider inflationary scenarios with positive potentials only: $V(\phi)>0$ during inflation.

    Using Eqs.~(\ref{EquHlo}) and (\ref{Equphilo}) and the effective potential $V_{eff}(\phi)$, we get the following equation~\cite{Pozdeeva:2020apf}:
\begin{equation}
\chi\simeq {}-2\frac{V}{U_0}{V_{eff}}_{,\phi}.
\label{EquphiloN}
\end{equation}

In terms of the effective potential, the slow-roll parameters are as follows:
\begin{equation}
\label{slrVeffe}
\varepsilon_1=\frac{V_{,\phi}}{U_0}{V_{eff}}_{,\phi}\,,\qquad
\delta_1= {}-\frac{2V^2}{3U_0^3}\xi_{,\phi}{V_{eff}}_{,\phi}\,.
\end{equation}
So, $|\varepsilon_1|\ll 1$ and $|\delta_1|\ll 1$ if the function ${V_{eff}}_{,\phi}$ is sufficiently small. It allows us to use the effective potential for construction of inflationary scenarios.

Also, we get
\begin{equation}
\label{eps2delta2phi}
\varepsilon_2(\phi)=\frac{U_0\delta_1}{2\xi_{,\phi}H^2\varepsilon_1}{\varepsilon_1}_{,\phi}\,,\qquad
\delta_2=\frac{U_0}{2H^2\xi_{,\phi}}{\delta_1}_{,\phi}.
\end{equation}
Explicit expressions of $\delta_2(\phi)$ and $\varepsilon_2(\phi)$ are given in Appendix~\ref{App}.

Using Eqs.~\eqref{r_slr} and \eqref{slrVeffe}, we get the tensor-to-scalar ratio
\begin{equation}
\label{rVeff}
r=16\frac{V^2}{U_0^3}\left({V_{eff}}_{,\phi}\right)^2.
\end{equation}

From Eqs.~\eqref{ns_slr} and \eqref{As_slr}, the spectral index $n_s$ and the scalar perturbation amplitude $A_s$ can be obtained~\cite{Pozdeeva:2020apf},
\begin{equation}
\label{nsVeff}
n_s=1+\frac{d}{dN}\ln\left(A_s\right)=1+\frac{2}{U_0}\left(2V {V_{eff}}_{,\phi\phi}+V_{,\phi} {V_{eff}}_{,\phi}\right),
\end{equation}
\begin{equation}
\label{As}
A_s\approx\frac{V}{6\pi^2 U_0^2\,r}=\frac{U_0}{96\pi^2 V\left({V_{eff}}_{,\phi}\right)^2}.
\end{equation}

Using Eq.~(\ref{EquphiloN}), we get how the e-folding number $N$ depends on $\phi$ in this approximation:
\begin{equation}
\label{N1}
\frac{dN}{d\phi}\simeq{}-\frac{U_0}{2V{V_{eff}}_{,\phi}}\,.
\end{equation}

\section{New slow-roll approximations}
\label{New-slow-roll-approximation}
~~~~Multiplying \eqref{Equ00delta1} to $H^2$ and substituting $\psi$ in terms of the slow-roll parameter $\delta_1$, we obtain:
\begin{equation}
\label{2indel1}
12\,{U_0} \left( 1-{\delta_1} \right)H^4 -2VH^2-\frac{\delta_1^2 U_0^2}{4\,\xi^2_{,\phi}}=0\,.
\end{equation}

    Equation \eqref{2indel1} is a quadratic equation in $H^2$. It always has a positive root if $\delta_1<1$:
\begin{equation}
\label{Qdelta1}
H^2=\frac{V}{12\,U_0
\, \left(1-\delta_1 \right)}+\frac{\sqrt{{V}^{2}\xi_{,\phi}^{2}+3\,U_0^3\delta_1^2\left(1-\delta_1\right)}}{12\,U_0
\, \left(1-\delta_1 \right)|\xi_{,\phi}|}\,.
\end{equation}

For searching more accurate approximations, it is reasonable to improve the expression for the Hubble parameter. Indeed, Eq.~(\ref{Equ0lo}) for the standard approximation does not ``feel'' the presence of the Gauss-Bonnet term ignoring all terms proportional to $\delta_1$. In the proposed approximations, the Hubble parameter depends on $\delta_1$ and the form of this dependence is the main difference between these new approximations.

\subsection{Approximation I}
~~~~Let us assume that $\delta_1\ll 1$ and expand the obtained expression to series with respect to the slow-roll parameter $\delta_1$:
\begin{equation}
\label{Qseries}
H^2\approx {\frac{V}{6\,U_0}}+{\frac{V}{6\,U_0}}\delta_1+{\cal{O}}(\delta^2_1)\,.
\end{equation}

    We construct a new slow-roll approximation keeping one more term in this expansion in comparison with the standard one:
\begin{equation}
\label{equ00slr}
H^2\simeq\frac{V}{6\,U_0}\left(1+\delta_1\right)=\frac{1}{6U_0^2}\left[U_0V+2V\xi_{,\phi}H\psi\right]\,.
\end{equation}

    In contrast to the standard slow-roll approximation, we do not neglect $\delta_1$, so, we should obtain $\delta_1(\phi)$ to get $H^2(\phi)$. To do it we use the following approximation of Eq.~(\ref{Equ11})  instead of Eq.~\eqref{EquHlo}:
\begin{equation}
\label{equ11slr}
\dot H \simeq{}-\frac{\psi^2+4H^3\xi_{,\phi}\psi}{4U_0\left(1-\delta_1\right)}.
\end{equation}

    Using Eq.~(\ref{delta}) and neglecting terms proportional to $\delta_1^3$, we rewrite~(\ref{equ11slr}) as follows
\begin{equation}
\label{equ11slrdel1}
\dot H\simeq{}-\frac{\delta_1}{4(1-\delta_1)}\left(\frac{U_0\delta_1}{4\xi_{,\phi}^2H^2}+2H^2\right)\simeq{}-\frac{H^2\delta_1}{2}-\frac{U_0\delta_1^2}{16\xi_{,\phi}^2H^2}-\frac{H^2\delta_1^2}{2}\,.
\end{equation}

    We neglect terms proportional to $\dot{\psi}$ and $\psi^2$ in Eq.~(\ref{Equphi}) and use Eq.~(\ref{delta}) to get the following approximate equation:
\begin{equation}
\label{equphislr1}
\frac{3U_0\delta_1}{2\xi_{,\phi}}={} -V_{,\phi} -12H^2\xi_{,\phi}\left(\dot{H}+H^2\right).
\end{equation}

    Substituting $H^2$ and $\dot H$ from Eqs.~(\ref{equ00slr}) and (\ref{equ11slrdel1}) into Eq.~\eqref{equphislr1} and neglecting terms, proportional to  $\delta_1^n$, where $n\geqslant 2$, we get
\begin{equation}
\delta_1(\phi)={}-\frac{2\, V^2\xi_{,\phi}\,{V_{eff}}_{,\phi}}{V^2\xi_{,\phi}^2+3\,U_0^3}\,.
\label{delta1phi}
\end{equation}

    The knowledge of $\delta_1(\phi)$ allows us to obtain $H^2(\phi)$ and $\chi(\phi)$. We obtain from Eq.~(\ref{equ00slr}) that
\begin{equation}
\label{H2slr}
H^2\simeq\frac{V}{6U_0}\left[1-\frac{2V^2\xi_{,\phi}{V_{eff}}_{,\phi}}{V^2\xi_{,\phi}^2+3U_0^3}\right]=\frac{V\left(9U_0^3-6U_0^2\xi_{,\phi}\,V_{,\phi}+\xi_{,\phi}^2V^2\right)}{18U_0\left(3U_0^3+\xi_{,\phi}^2V^2\right)}.
\end{equation}

    Therefore, Eq.~(\ref{delta}) gives
\begin{equation}
\label{apprIequdphidN}
\chi=\frac{U_0\delta_1}{2\xi_{,\phi}H^2}\simeq{}-\frac{6U_0^2V{V_{eff}}_{,\phi}}{V^2\xi_{,\phi}^2+3U_0^3-2V^2\xi_{,\phi}{V_{eff}}_{,\phi}}
={}-\frac{6U_0^2\left(3U_0^2V_{,\phi}+\xi_{,\phi}V^2\right)}{V\left(9U_0^3-6U_0^2\xi_{,\phi}V_{,\phi}+\xi_{,\phi}^2V^2\right)}\,,
\end{equation}
hence,
\begin{equation}
\label{apprIdNDdphi}
\frac{dN}{d\phi}= {}-\frac{V^2\xi_{,\phi}^2+3U_0^3-2V^2\xi_{,\phi}{V_{eff}}_{,\phi}}{6U_0^2V{V_{eff}}_{,\phi}}\,.
\end{equation}

    Using Eq.~(\ref{H2slr}), we get the slow-roll parameters as functions of $\phi$:
\begin{equation}
\label{apprIeps1phi}
\varepsilon_1(\phi)={}-\frac{d\phi}{dN}\frac{d\ln(H)}{d\phi}=\frac{6U_0^2\left(3U_0^2V_{,\phi}+\xi_{,\phi}V^2\right)}{V\left(9U_0^3-6U_0^2\xi_{,\phi}V_{,\phi}+\xi_{,\phi}^2V^2\right)}\,\frac{d\ln(H)}{d\phi}\,,
\end{equation}
where
\begin{equation*}
\frac{d\ln(H)}{d\phi}=\frac{V_{,\phi}}{2V}+\frac{
\xi_{,\phi}\xi_{,\phi\phi}V^2+\xi_{,\phi}^2VV_{,\phi}-3U_0^2\xi_{,\phi\phi}V_{,\phi}-3U_0^2\xi_{,\phi}V_{,\phi\phi}}
{9U_0^3-6U_0^2\xi_{,\phi}V_{,\phi}+\xi_{,\phi}^2V^2}-\frac{\xi_{,\phi}\xi_{,\phi\phi}V^2+\xi_{,\phi}^2VV_{,\phi}}{3U_0^3+\xi_{,\phi}^2V^2}\!.
\end{equation*}

 Explicit expressions of parameters $\varepsilon_2(\phi)$ and $\delta_2(\phi)$ obtained by Eq.~(\ref{eps2delta2phi}) are given in Appendix~\ref{App}.

 The knowledge of the slow-roll parameters as functions of $\phi$ allows us to get the inflationary parameters $n_s(\phi)$, $r(\phi)$, and $A_s(\phi)$ by formulae (\ref{ns_slr})-(\ref{As_slr}).

\subsection{Approximation II}
~~~~Another way of generalizing Eq.~(\ref{Equ0lo}) is the following one. We neglect the term proportional to $\delta_1^2$ in Eq.~(\ref{2indel1}) and get a nonzero solution:
\begin{equation}
\label{apprIIH2}
H^2=\frac{V}{6U_0(1-\delta_1)}.
\end{equation}

If we expand this expression into powers of $\delta_1$ and keep two first terms, then we return to the approximation~I. Thus, we can expect that Eq.~(\ref{apprIIH2}) gives better accuracy. Clearly, these both approximations are not valid at $\delta_1=1$. Note, however, that in the approximation~I, Eq.~(\ref{equ00slr}) shows that  the value of $H^2$ at $\delta_1=1$ is two times more than the corresponding value at $\delta_1=0$, whereas in the approximation~II, Eq.~(\ref{apprIIH2}) gives infinite value of the Hubble parameter at $\delta_1=1$.
So, the approximation~I may be more accurate than the approximation~II for $\delta_1$ close to $1$ in some cases. That is why we consider both approximations on the concrete examples and compare them with numerical calculations.

Using Eqs.~(\ref{delta}) and \eqref{apprIIH2}, we get
\begin{equation}
\label{apprIIdH2dN}
\frac{\,dH^2}{dN}=\frac{V_{,\phi}\,\delta_1}{12\xi_{,\phi}H^2(1-\delta_1)}+\frac{V\delta_1\delta_2}{6U_0(1-\delta_1)^2}=\frac{U_0\,V_{,\phi}\,\delta_1}{2\xi_{,\phi}V}+\frac{V\delta_1\delta_2}{6U_0(1-\delta_1)^2}
\end{equation}
and
\begin{equation}
\label{epsilon1fromDiff}
\varepsilon_1={}-\frac{3\,U^2_0\,V_{,\phi}}{2 V^2\xi_{,\phi}}\delta_1(1-\delta_1)-\frac{\delta_1\delta_2}{2\,(1-\delta_1)}\,.
\end{equation}

    From \eqref{delta2}, we get
\begin{equation}
\label{del2}
\dot{\psi}\approx\frac{U_0\delta_1}{2\xi_\phi}\left(\delta_2+\varepsilon_1-\frac{3U_0^2\xi_{,\phi\phi}\delta_1}{V\,\xi_{,\phi}^2}\right)\,.
\end{equation}

    Substituting formulae \eqref{apprIIH2}, \eqref{epsilon1fromDiff}, and \eqref{del2} into \eqref{Equphi}, we get
\begin{equation}
\frac{U_0\delta_1}{2\xi_\phi}\left(\delta_2+\varepsilon_1-\frac{3U_0^2\xi_{,\phi\phi}\delta_1}{V\,\xi_{,\phi}^2}\right)
+\frac{3U_0\delta_1}{2\xi_{,\phi}}+V_{,\phi}+\frac{12\xi_{,\phi}V^2\,V_{,\phi}}{36\,U_0^2(1-\delta_1)^2}(1-\varepsilon_1)=0\,.
\end{equation}

    Multiplying this equation  to $(1-\delta_1)^2$ and supposing that any products of slow-roll parameters are negligible, we obtain the following linear equation in $\delta_1$:
\begin{equation}
\frac{9}{2} \left(\frac{U_0^3}{\xi_{,\phi}}-V_{,\phi}\,U_0^2\right) {\delta_1}
+3\,V_{,\phi}\,U_0^2+V^2\xi_{,\phi}=0.
\end{equation}

    Solving this equation, we get
\begin{equation}
\label{apprIIdel1phi}
\delta_1(\phi)={}-\frac{2\xi_{,\phi}\left(3U_0^2V_{,\phi}+V^2\xi_{,\phi}\right)}{9U_0^2\left(U_0-\xi_{,\phi}V_{,\phi}\right)}.
\end{equation}

Now we can express $H^2$, $\chi$, $N_{,\phi}\,$, and $\varepsilon_1$ via $\phi$:
\begin{equation}
\label{apprIIH2phi}
H^2(\phi)\simeq\frac{V}{6U_0(1-\delta_1(\phi))}=\frac{3{U_0}V(U_0-\xi_{,\phi}V_{,\phi})}{2(9U_0^3-3U_0^2\xi_{,\phi}V_{,\phi}+2\xi_{,\phi}^2V^2)},
\end{equation}
\begin{equation}
\label{apprIIequdphidN}
\chi=\frac{U_0\delta_1}{2\xi_{,\phi}H^2}\simeq{}-\frac{2\left(3U_0^2V_{,\phi}+\xi_{,\phi}V^2\right)
\left(9U_0^3-3U_0^2\xi_{,\phi}V_{,\phi}+2\xi_{,\phi}^2V^2\right)}{27U_0^2V{\left(U_0-\xi_{,\phi}V_{,\phi}\right)}^2},
\end{equation}
\begin{equation}
\label{apprIIequdNdphi}
\frac{dN}{d\phi}=\frac{1}{\chi}={}-\frac{27U_0^2V{\left(U_0-\xi_{,\phi}V_{,\phi}\right)}^2}{2\left(3U_0^2V_{,\phi}+\xi_{,\phi}V^2\right)\left(9U_0^3-3U_0^2\xi_{,\phi}V_{,\phi}+2\xi_{,\phi}^2V^2\right)},
\end{equation}
\begin{equation}
\label{apprIIeps1phi}
\varepsilon_1(\phi)=\frac{\left(3U_0^2V_{,\phi}+\xi_{,\phi}V^2\right)\left(9U_0^3-3U_0^2\xi_{,\phi}V_{,\phi}+2\xi_{,\phi}^2V^2\right)}
{27U_0^2V{\left(U_0-\xi_{,\phi}V_{,\phi}\right)}^2}\,\frac{d\ln(H^2)}{d\phi}\,,
\end{equation}
where
\begin{equation*}
\label{apprIIdH2dphi}
\frac{d\ln(H^2)}{d\phi}=\frac{V_{,\phi}}{V}+\frac{\xi_{,\phi\phi}V_{,\phi}+\xi_{,\phi}V_{,\phi\phi}}{\xi_{,\phi}V_{,\phi}-U_0}+\frac{3U_0^2\xi_{,\phi\phi}V_{,\phi}+3U_0^2\xi_{,\phi}V_{,\phi\phi}-4\xi_{,\phi}\xi_{,\phi\phi}V^2
-4\xi_{,\phi}^2VV_{,\phi}}{9U_0^3-3U_0^2\xi_{,\phi}V_{,\phi}+2\xi_{,\phi}^2V^2}\,.
\end{equation*}

    Using the obtained expressions of $\delta_1(\phi)$ and $\varepsilon_1(\phi)$ and Eq.~\eqref{eps2delta2phi}, we calculate parameters $\varepsilon_2(\phi)$ and $\delta_2(\phi)$. Explicit expressions of these parameters are given in Appendix~\ref{App}.
After this, one can obtain $n_s(\phi)$, $r(\phi)$, and $A_s(\phi)$  due to formulae (\ref{ns_slr})-(\ref{As_slr}).

\section{Inflationary models with monomial potentials}

\subsection{The choice of the function $\xi(\phi)$}
~~~~Inflationary models with minimally coupled scalar field and quadratic and quartic potentials, proposed by Linde in Refs.~\cite{Linde:1981mu,Linde:1983gd}, have been ruled out observationally~\cite{Planck:2018jri}. One of the way to construct valuable inflationary models with even monomial potentials is adding of the Gauss-Bonnet term multiplying on some function of the scalar field. Models with the potential $V=V_0\phi^n$, where $n=2$ or $n=4$ and the function $\xi$ being inversely proportional to $V$ have been proposed in Ref.~\cite{Guo:2010jr}, using the standard slow-roll approximation. The obtained inflationary parameters did not contradict to the observation data of that time, but numerical calculations in the case of $n=4$ indicated that the slow-roll parameter $\varepsilon_1$ never exceeds unity, leading thus to eternal inflation~\cite{vandeBruck:2015gjd}. In order to have the exit from inflation, we modify the coupling function so that
\begin{equation}
\label{xiphi}
\xi=\frac{CU_0^2}{V+\Lambda},
\end{equation}
where $C$ and $\Lambda$ are positive constants. Apart from reproducing the desired cosmological evolution, such a modification is natural in general, removing a singular behavior at $\phi=0$. This modification gives us an ultimate exit from inflation when $\phi$ becomes small enough.

    We assume that the initial value of the scalar field is positive, and the scalar field tends to zero during inflation. Calculating the derivative of the effective potential~(\ref{Veff}),
\begin{equation}
\label{DVeff}
{V_{eff}}_{,\phi}=\frac{U_0^2n\left(V_0^2
\left(3-C\right)\phi^{2\,n}+6\,\Lambda\,V_0{\phi}^{n}+3\,{\Lambda}^{2}
\right)}{3V_0{\phi}^{n+1}\left(V_0\,{\phi}^{n}+\Lambda\right)^{2}}\,,
\end{equation}
we find that ${V_{eff}}_{,\phi}>0$ for any $\phi>0$ at $C<3$. It is a sufficient condition~\cite{Pozdeeva:2019agu} that a de~Sitter solution does not exist at any $\phi>0$. This condition allows us to get an inflationary model without any fine-tuning of the initial data.

\subsection{Models with quadratic and quartic potentials}
~~~~To construct inflationary models in the cases of $n=2$ and $n=4$ we solve numerically system~\eqref{DynSYSN} and calculate slow-roll parameters by formulae~\eqref{epsilon} and \eqref{delta}. After this, we calculate inflationary parameters using Eqs.~(\ref{ns_slr})--(\ref{As_slr}). We choose such values of parameters of the considered models that the corresponding inflationary parameters do not contradict to recent observation data. We compare different slow-roll approximations for the considered inflationary models with the chosen values of parameters.

    For the model with the potential $V=V_0\,\phi^2$ and the following values of parameters:
\begin{equation}
\label{Paramphi2}
U_0=\frac{M_\mathrm{Pl}^2}{2}\,, \quad C=2.754,\quad V_0= 4.05\times10^{-11}M_\mathrm{Pl}^2,\quad \Lambda = 1.0125\times 10^{-12}M_\mathrm{Pl}^4\,,
\end{equation}
numerical integration of system \eqref{DynSYSN} gives the following values of the inflationary parameters:
\begin{equation}
\label{InflParamphi2}
A_s={2.0968}\times 10^{-9}\,,\qquad n_s=0.9654,\qquad r=0.0102.
\end{equation}

\begin{figure}
\includegraphics[scale=0.41]
{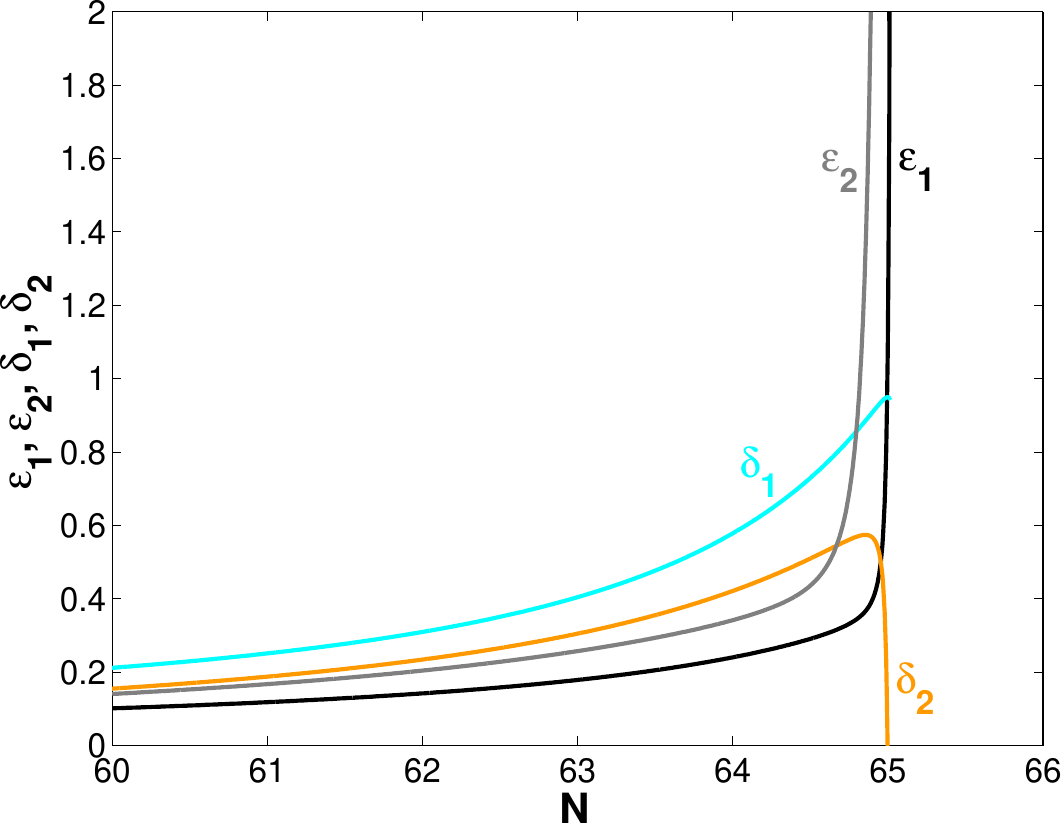}\quad
\includegraphics[scale=0.41]{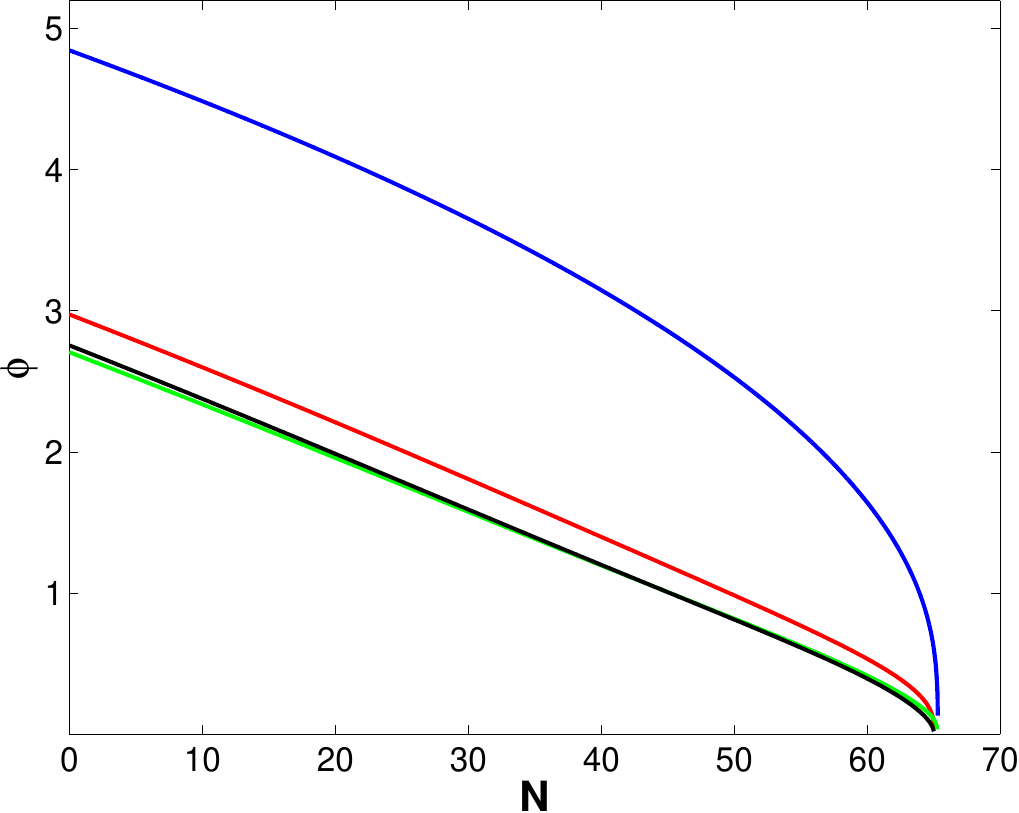}
\caption{The inflationary model with $V(\phi)=V_0\phi^2$.
Slow-roll parameters as functions of $N$ at the end of inflation found by numerical integration of equations of motion without any approximation are presented in the left panel. Values of the function $\phi(N)$ in units of $M_\mathrm{Pl}$ obtained numerically or using slow-roll approximations are presented in the right panel. The black line is the result of the numerical integration of system~(\ref{DynSYS}). The blue curve is obtained in the standard approximation using Eq.~(\ref{EquphiloN}), red --- in the approximation~I using Eq.~(\ref{apprIequdphidN}), green --- in the approximation~II by Eq.~(\ref{apprIIequdphidN}).
The initial values $\phi(0)=\phi_0$ are given in Table~\ref{TablePhi2a}.  }
\label{Fig1phi2}
\end{figure}

    The inflationary parameters are calculated at $\phi_0=2.7565
    M_\mathrm{Pl}$ that corresponds to $N=0$. The inflation finishes at $N_{end}=65$ (see Fig.~\ref{Fig1phi2}) that corresponds to $\phi_{end}=0.0294 M_\mathrm{Pl}$. So, the constructed inflationary scenario does not contradict to the observation data~(\ref{Inflparamobserv}).

     Let us check the possibility to get this inflationary model using slow-roll approximations. Numerical calculations show that the slow-roll parameters are less than unity almost up to the end of inflation, and the field $\phi$ is a monotonic function (see Fig.~\ref{Fig1phi2}). So, the necessary conditions for the use of slow-roll approximations are satisfied.

\begin{figure}
\includegraphics[scale=0.41]{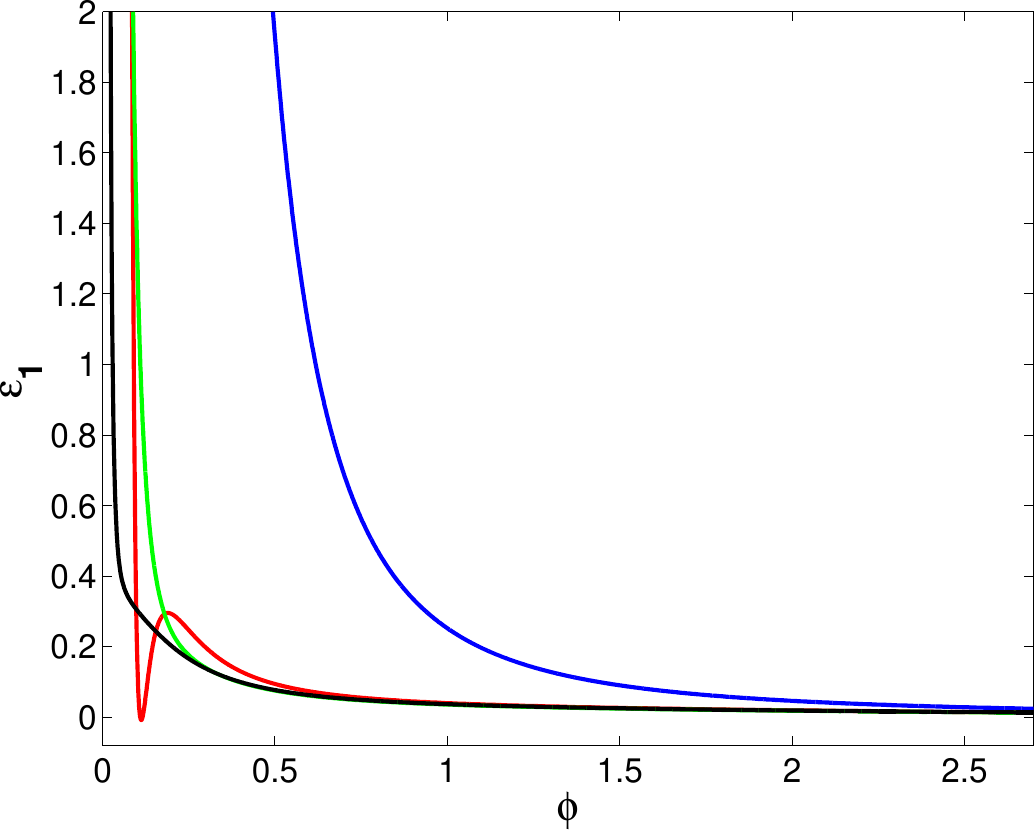}\qquad
\includegraphics[scale=0.41]{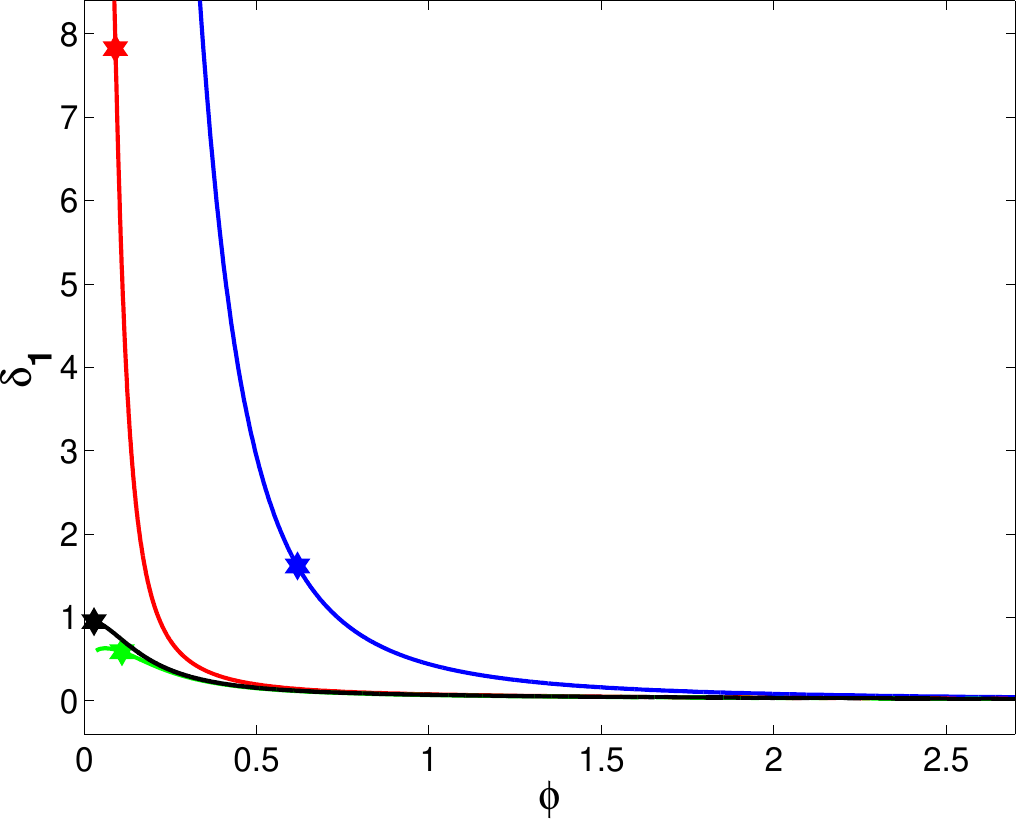}
\caption{The slow-roll parameters $\varepsilon_1(\phi)$ (left panel) and $\delta_1(\phi)$ (right panel) for the model with $V(\phi)=V_0\phi^2$. The black line is the result of the numerical integration of the system~(\ref{DynSYS}), blue curves are obtained in the standard approximation, red curves in the approximation~I, and green curves in the approximation~II. The stars denote the end of the inflation (when $\varepsilon_1=1$). Values of $\phi$ are given in units of $M_\mathrm{Pl}$.}
\label{Fig2phi2}
\end{figure}

    In Figs.~\ref{Fig1phi2} and~\ref{Fig2phi2}, we can see that the standard approximation deviates significantly from the numerical evolution. The behavior of the slow-roll parameters $\varepsilon_1(\phi)$ and $\delta_1(\phi)$ at the end of inflation is presented in Fig.~\ref{Fig2phi2}. One can see that the values of $\varepsilon_1(\phi)$ are essentially closer to the numerical results in both new approximations than in the standard approximation. It indicates the end of inflation at a much higher value of $\phi_{end}$, defined by condition $\varepsilon_1(\phi_{end})=1$ in the standard approximation, than the numerical calculations show (see Fig.~\ref{Fig2phi2}). It is not surprising, since the standard approximation predicts the end of inflation even in the absence of the cut-off parameter $\Lambda$ in the coupling function. On the contrary, both more involved approximations indicate the end of inflation close to the corresponding numerical value. The behavior of $\delta_1(\phi)$ is essentially better in the approximation~II than in the approximation~I.

    We use two methods in order to compare numerical results with approximations for large values of $\phi$. First, we fix the number of e-foldings to be $N=65$ and find corresponding values of $\phi_{end}$ and $\phi_0$ in all approximations. In the left panel of Fig.~\ref{Fig1phi2}, one can see that the value of $\phi_0=\phi(0)$ in the standard approximation overestimates significantly the corresponding values of scalar field obtained numerically or in the proposed approximations. The inflationary parameters calculated by approximate formulae are presented in Table~\ref{TablePhi2a}. We see that the value of $n_s$ and $r$ are suitable in all approximation, whereas the approximate values of $A_s$ are beyond the currently acceptable region.

\begin{figure}[hbtp]
\includegraphics[scale=0.41]{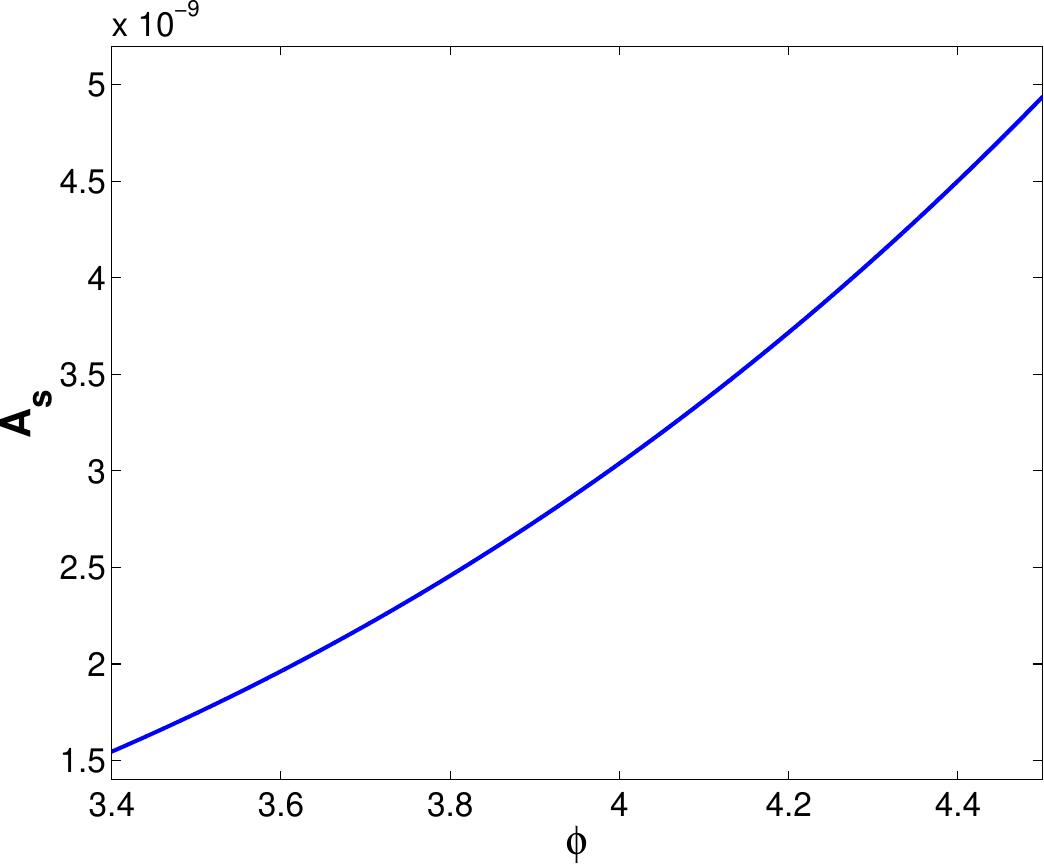}\quad
\includegraphics[scale=0.41]{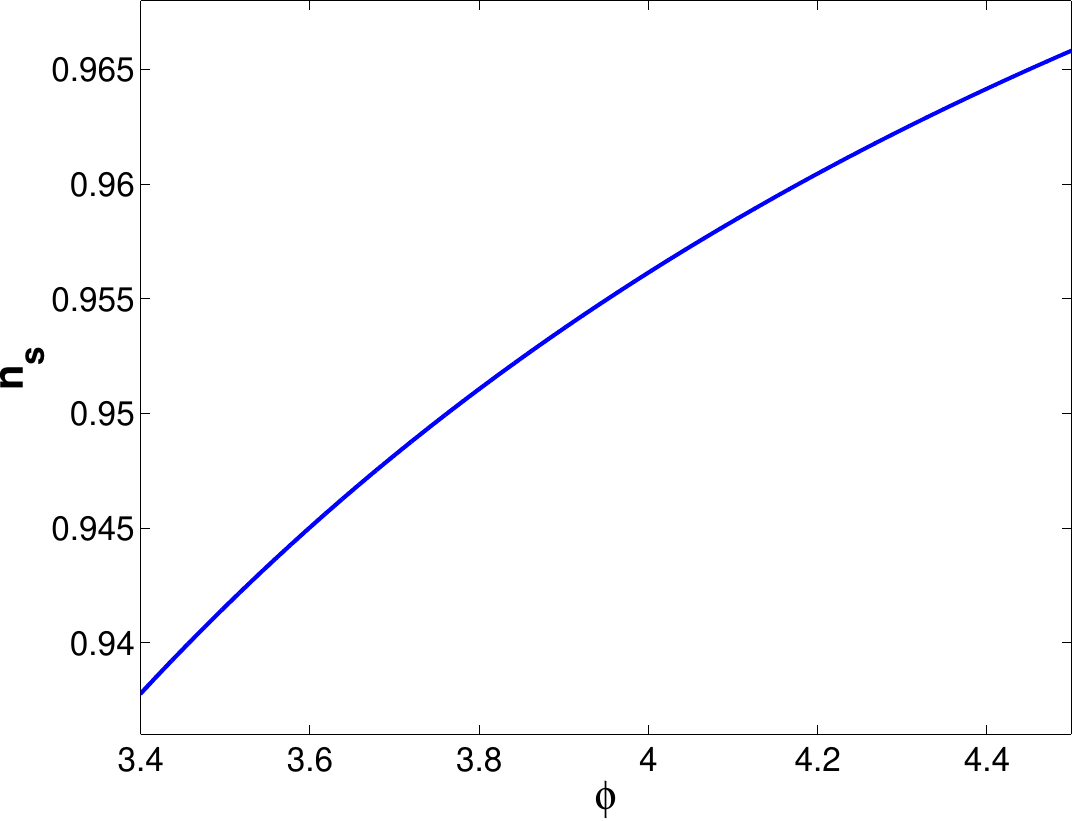}
\caption{The inflationary parameters $A_s$ and $n_s$ as functions of the scalar field $\phi$ for the model with $V(\phi)=V_0\phi^2$ in the standard approximation. Values of $\phi$ are given in units of $M_\mathrm{Pl}$.}
\label{Fig3phi2}
\end{figure}

    Let us check the possibility to get the suitable values of inflationary parameters. As the number of e-folding is not known exactly, we can use the following method. We solve the equation $A_s(\phi_{in})=2.1\times10^{-9}$, get $\phi_{in}$ and calculate values of $n_s(\phi_{in})$ and $r(\phi_{in})$ in all approximations. We also get the corresponding e-folding numbers from $N(\phi_{in})$ up to the end of inflation. The results are presented in Table~\ref{TablePhi2b}. One can see that the standard approximation seriously underestimates the number of e-foldings. The index $n_s$ is beyond allowed region as well. Looking at Fig.~\ref{Fig3phi2}, one can see that in the standard approximation, either the value of $A_s(\phi_{in})$ is too large or the value of $n_s$ is too small for the model considered even if we do not impose any restrictions on the e-folding number. So, we can conclude that the considered inflationary model cannot be found using the standard approximation. On the other hand, it can be found using any of two approximations proposed in this paper, because values of the inflationary parameters belong to the currently acceptable region (see Table~\ref{TablePhi2b}).

\begin{table}
\caption{\label{TablePhi2a} \textbf{Numerical and approximate values of parameters, characterizing the inflationary dynamic in the model with the quadratic potential.}}
\vspace{4mm}
\begin{tabular}{|c|c|c|c|c|}
\hline
\bf{Parameter} & \bf{Numerical} & \bf{Standard} & \bf{Approximation I} & \bf{Approximation II}\\
{} &  \bf{results} & \bf{approximation} & {\ } & { \ }\\
\hline
$\phi_0/M_\mathrm{Pl}$ & {$2.7565$} & $4.8472$ & $2.9757$ & $2.7082$\\
\hline
$10^9A_s(\phi_0)$ & $2.097$ & $6.696$ & {$2.491$} & {$1.985$}\\
\hline
$n_s(\phi_0)$ & $0.965$ & {$0.971$} & {$0.967$} & {$0.965$}\\
\hline
$r(\phi_0)$ & $0.0102$ & $0.0096$ & {$ 0.0099$} & {$0.0104$}\\
\hline
$\phi_{end}/M_\mathrm{Pl}$ & $0.0294$ & $0.6184$ & $0.0906$ & $0.1097$\\
\hline
$\delta_1(\phi_{end})$ & {$ 0.950$} & $1.62$ & $7.82$ & {$0.590$}\\
\hline
$N(\phi_{end})$ & $65.0$ & $65.0$ & $65.0$ & $65.0$\\
\hline
\end{tabular}
\end{table}

\begin{table}
\caption{\label{TablePhi2b} \textbf{Values of the inflationary parameters for the model with the quadratic potential in different approximations.}}
\vspace{4mm}
\begin{tabular}{|c|c|c|c|}
\hline
\bf{Parameter} & \bf{Standard} & \bf{Approximation I} & \bf{Approximation II}\\
{} &  \bf{approximation} & {} & {}\\
\hline
$\phi_{in}/M_\mathrm{Pl}$ & $3.6589$ & $2.7912$ & $2.7676$\\
 \hline
$10^9A_s(\phi_{in})$ & $2.10$ & $2.10$ & $2.10$\\
\hline
$n_s(\phi_{in})$ & $0.947$ & $0.965$ & $0.966$\\
\hline
$r(\phi_{in})$ & $0.0174$ &  $0.0104$ & $0.0102$\\
\hline
$N(\phi_{end})-N(\phi_{in})$ &  $35.1$ & $60.0$ &  $66.6$\\
\hline
\end{tabular}
\end{table}


 The standard approximation gives the behavior of $\phi(N)$ essentially different from the numerical results without any approximation (see Fig.~\ref{Fig1phi2}). By this reason, the standard approximation predicts the exit from inflation to appear considerably earlier than for the exact results. So, any physical quantity explicitly depending on the dynamics near end of inflation (note that observable perturbations have been born well before the end of inflation and their dependence upon the details of the exit from inflation is indirect through the dependence of the scalar field upon the number of e-foldings), if calculated via standard approximation, would give very inaccurate results. We illustrate this by calculating the speed of the perturbation propagations.

\begin{figure}
\includegraphics[scale=0.41]{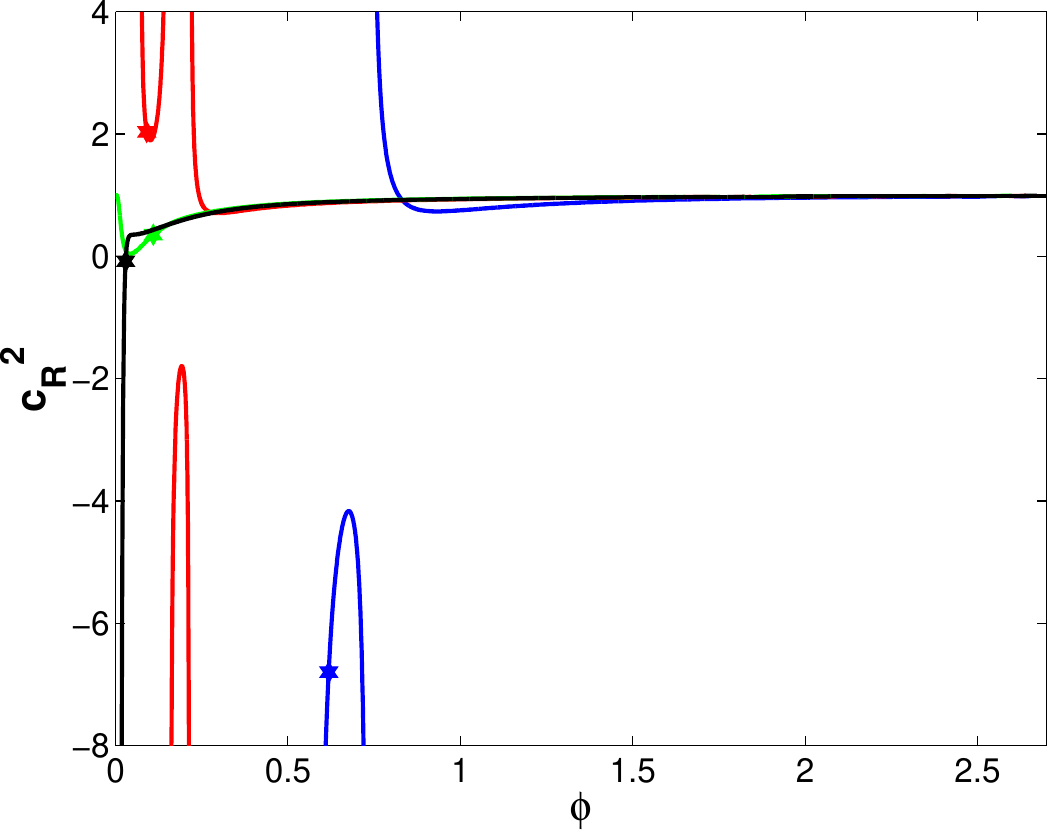}\quad
\includegraphics[scale=0.41]{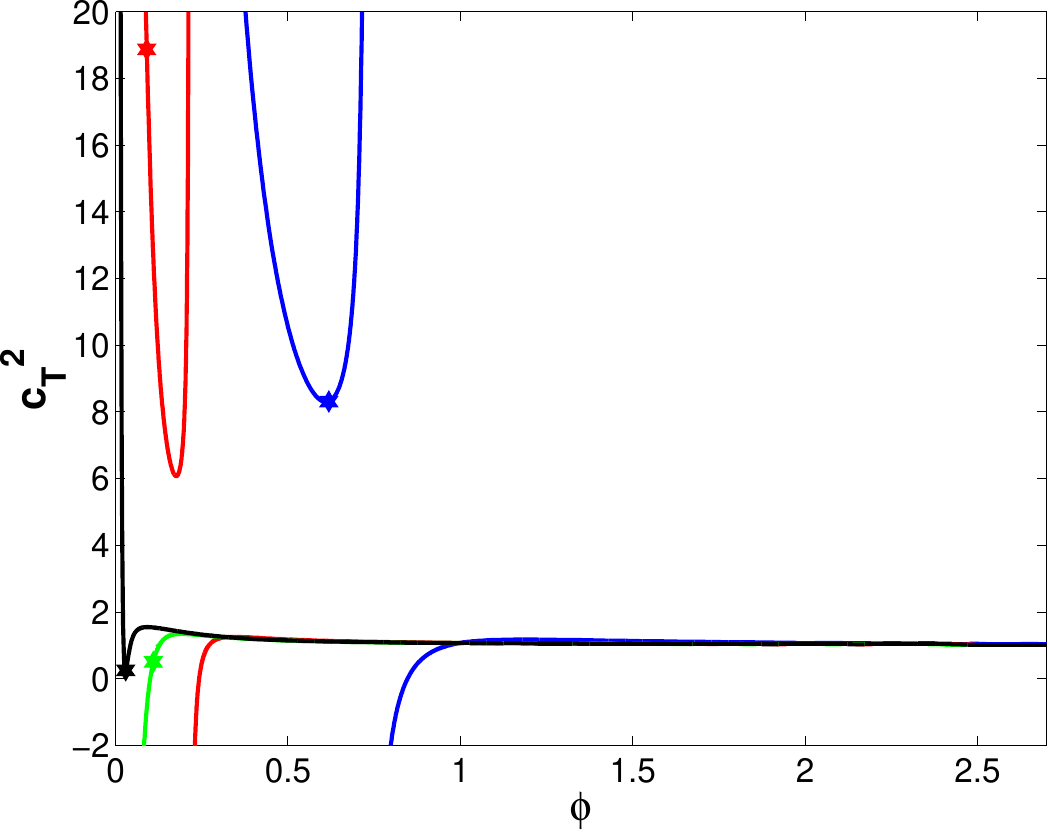}
\caption{The scalar perturbation propagation wave speed and the gravitational wave propagation speed for the model with $V(\phi)=V_0\phi^2$. The black line is the result of the numerical integration of the system~(\ref{DynSYS}), blue curves are obtained in the standard approximation, red curves in the approximation~I, and green curves in the approximation~II. The stars denote the end of the inflation (when $\varepsilon_1=1$). Values of $\phi$ are given in units of $M_\mathrm{Pl}$.}
\label{Fign2speed}
\end{figure}

Using Eqs.~(\ref{cR})--(\ref{cTepsdel}), we compare exact and approximate values of the scalar perturbation propagation wave speed $c_R$ and of the gravitational wave propagation speed $c_T$. Figure \ref{Fign2speed} shows that new approximations give essentially more accurate values of $c_R^2$ and $c_T^2$ than the standard approximation. Note that the approximation~II gives better results than the approximation~I at the end of inflation.

    The situation is similar for the model with the fourth-order potential $V=V_0\phi^4$. For parameters
\begin{equation}
V_0 = 3.4\times10^{-11},\quad C = 2.856,\quad \Lambda = 5.95\times10^{-13} M_\mathrm{Pl}^4\,,
\end{equation}
numerical calculations show that the inflationary scenario does not contradict the current observation data and the e-folding number $N(\phi_{end})=60.6$. We get unappropriated results in the standard approximation, whereas, new approximations, as in the previous example, work essentially better (see Table~\ref{TablePhi4a}).

\begin{table}
\caption{\label{TablePhi4a} \textbf{Numerical and approximate values of parameters, characterizing the inflationary dynamic in the model with the quartic potential.}}
\vspace{4mm}
\begin{tabular}{|c|c|c|c|c|}
\hline
\bf{Parameter} & \bf{Numerical} & \bf{Standard} & \bf{Approximation I} & \bf{Approximation II}\\
{ } & \bf{results} & \bf{approximation} & { } & { }\\
\hline
$\phi_0/M_\mathrm{Pl}$ & {$1.4019$} & {$4.9705$} & {$1.4898$} & $1.3974$\\
\hline
$10^9A_s(\phi_0)$ & {$2.096$} & {$117.2$} & {$2.599$} & {$2.017$}\\
\hline
$n_s(\phi_0)$ & {$0.965$} & {$ 0.953$} & {$0.965$} & {$0.965$}\\
\hline
$r(\phi_0)$ & $0.0044$ & $0.0120$ & $0.0045$ & $0.0045$\\
\hline
$\phi_{end}/M_\mathrm{Pl}$ & $0.2017$ & $0.8899$ & $0.3048$ & $0.3037$\\
\hline
$\delta_1(\phi_{end})$ & $0.885$ & $1.80$ & $4.23$ & $0.577$\\
\hline
$N(\phi_{end})$ & $60.6$ & $60.6$ & $60.6$ & $60.6$\\
\hline
\end{tabular}
\end{table}

\begin{figure}
\includegraphics[scale=0.41]{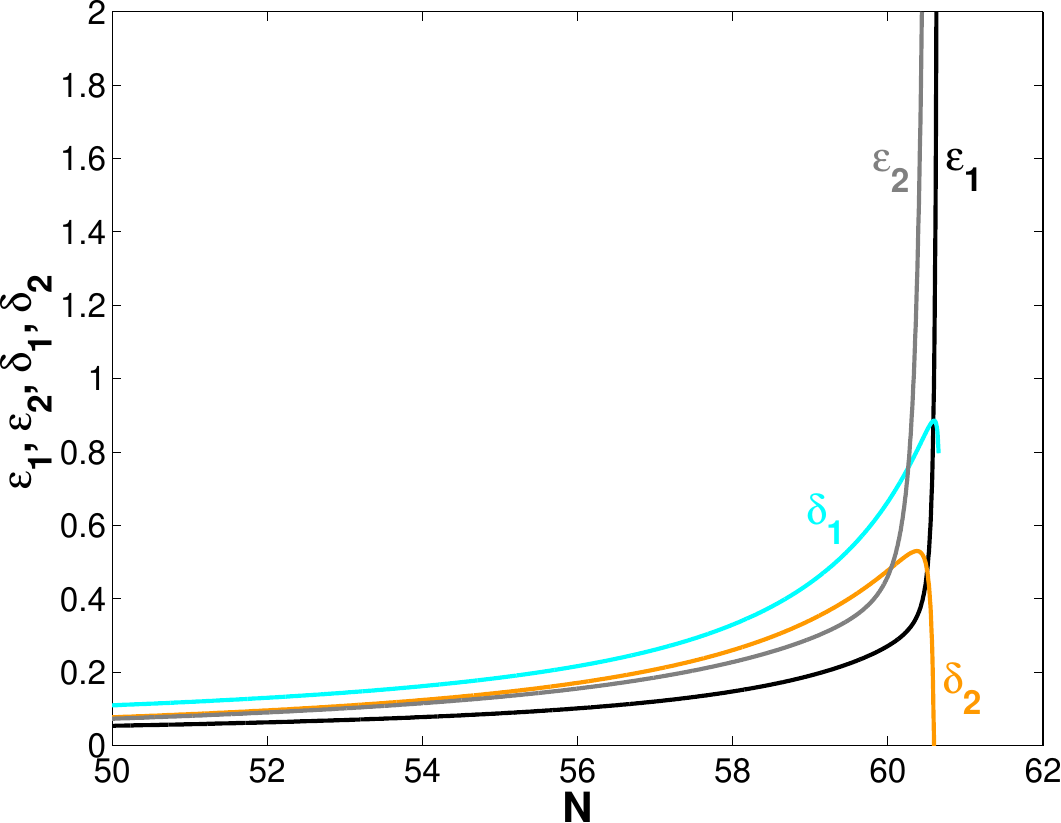}\quad
\includegraphics[scale=0.41]{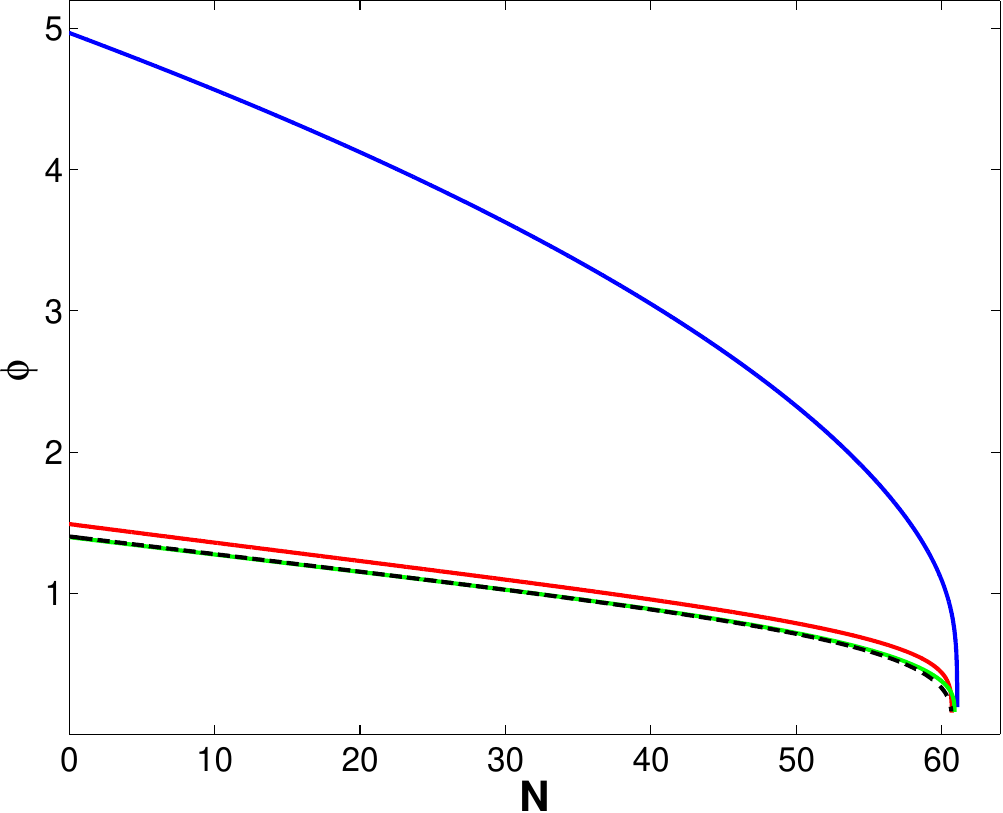}
\caption{The inflationary model with $V(\phi)=V_0\phi^4$.
Slow-roll parameters as functions of $N$ at the end of inflation found by numerical integration of equations of motion without any approximation are presented in the left panel. Values of the function $\phi(N)$ in units of $M_\mathrm{Pl}$ obtained numerically or using slow-roll approximations are presented in the right panel. The black line is the result of the numerical integration of system~(\ref{DynSYS}). The blue curve is obtained in the standard approximation using Eq.~(\ref{EquphiloN}), red --- in the approximation~I using Eq.~(\ref{apprIequdphidN}), green --- in the approximation~II by Eq.~(\ref{apprIIequdphidN}).
The initial values $\phi(0)=\phi_0$ are given in Table~\ref{TablePhi4a}.}
\label{Fig1phi4}
\end{figure}

\begin{figure}
\includegraphics[scale=0.41]{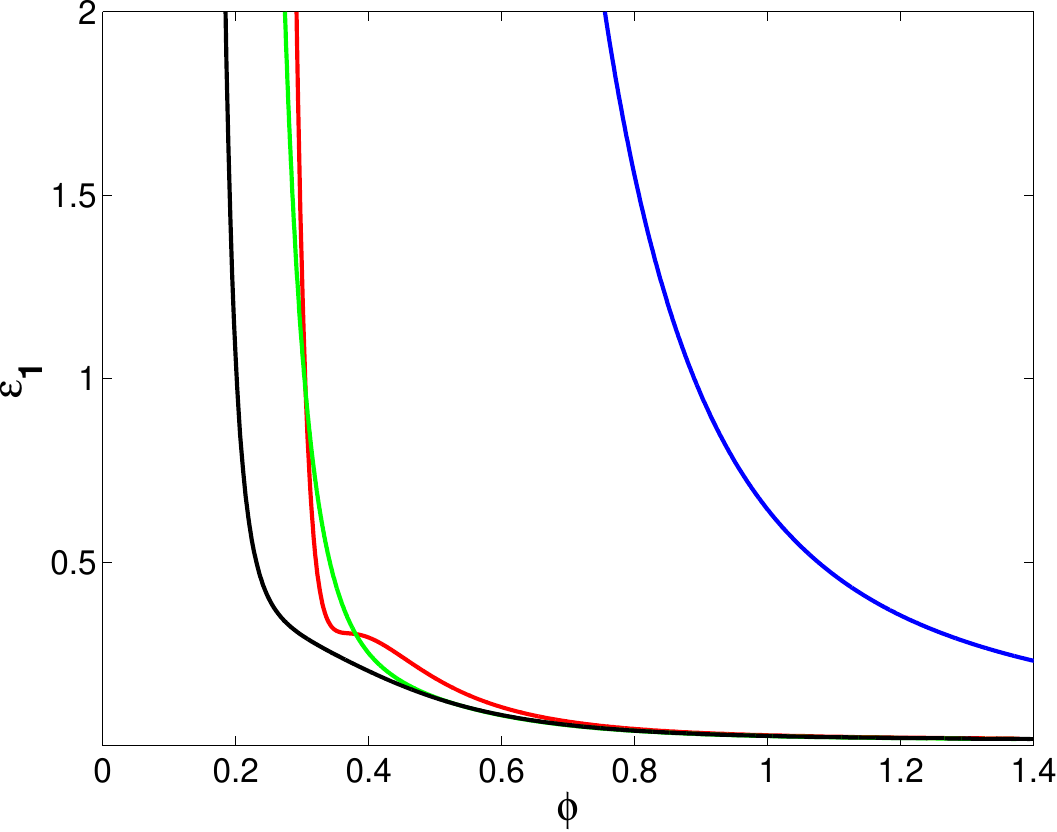}\quad
\includegraphics[scale=0.41]{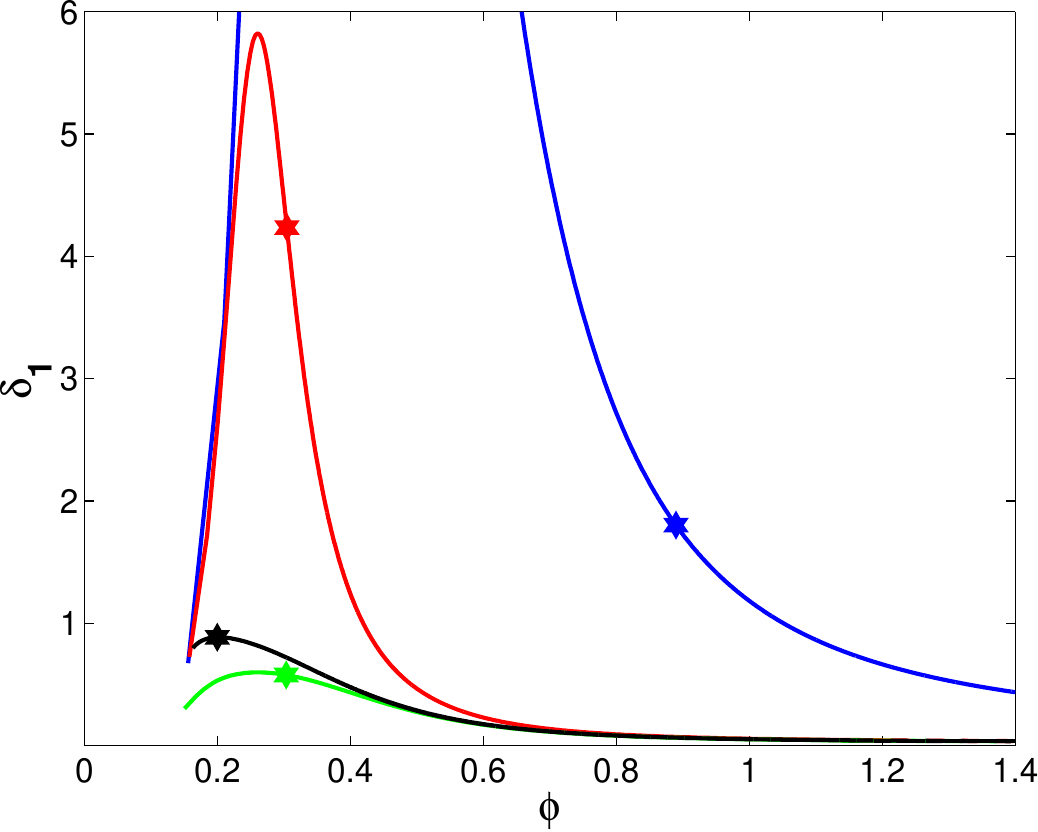}
\caption{The slow-roll parameters $\varepsilon_1(\phi)$ (left panel) and $\delta_1(\phi)$ (right panel) for the model with $V(\phi)=V_0\phi^4$. The black line is the result of the numerical integration of the system~(\ref{DynSYS}), blue curves are obtained in the standard approximation, red curves --- in the approximation~I, and green curves --- in the approximation~II. The stars denote the end of the inflation (when $\varepsilon_1=1$). Values of $\phi$ are given in units of $M_\mathrm{Pl}$.}
\label{Fig2phi4}
\end{figure}

\begin{figure}
\includegraphics[scale=0.41]{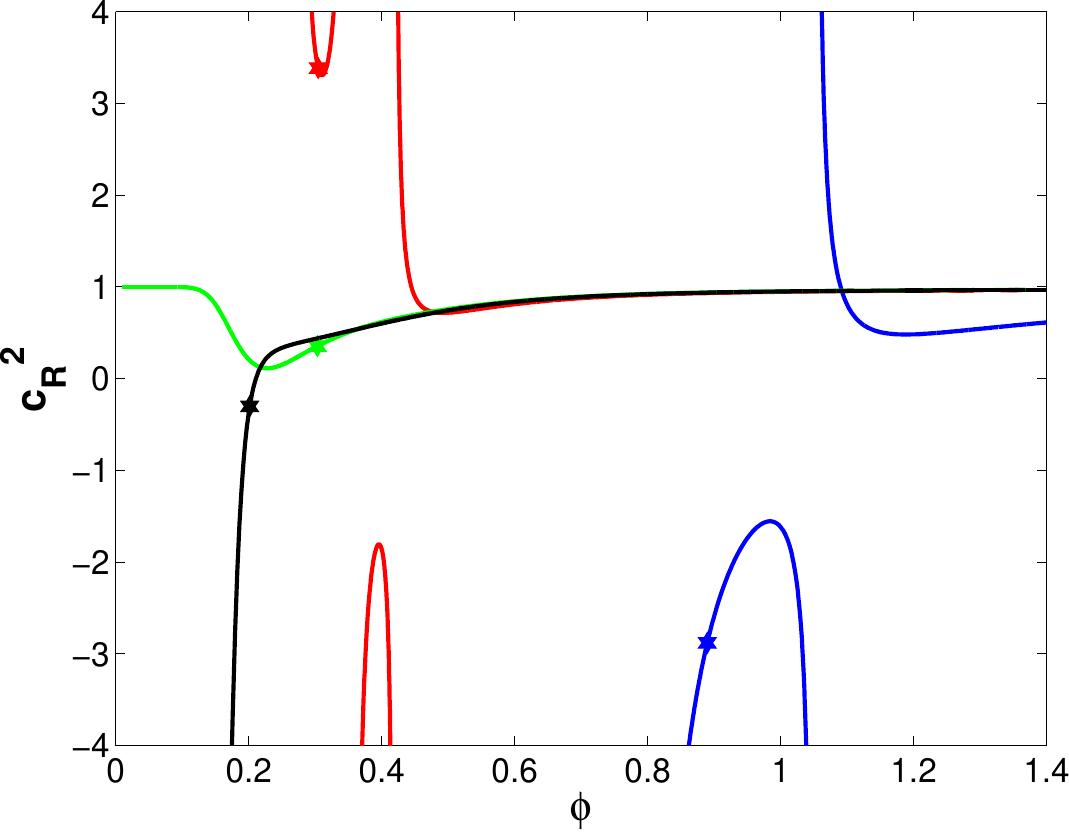}\quad
\includegraphics[scale=0.41]{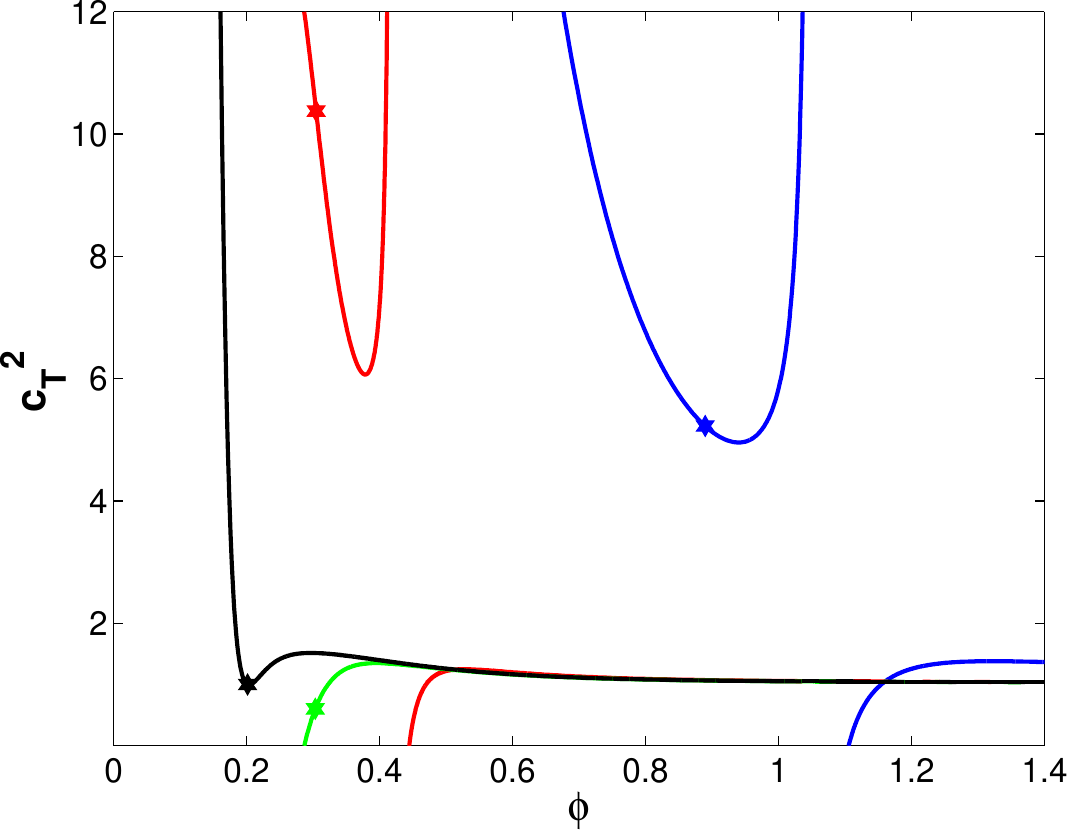}
\caption{The scalar perturbation propagation wave speed and the gravitational wave propagation speed for the model with $V(\phi)=V_0\phi^4$. The black line is the result of the numerical integration of the system~(\ref{DynSYS}), blue curves are obtained in the standard approximation, red curves in the approximation~I, and green curves in the approximation~II. The stars denote the end of the inflation (when $\varepsilon_1=1$). Values of $\phi$ are given in units of $M_\mathrm{Pl}$.}
\label{Fign4speed}
\end{figure}

New approximations give essentially more accurate behavior of the scalar field during inflation than the standard approximation (see Fig.~\ref{Fig1phi4}). In Figs.~\ref{Fig2phi4} and \ref{Fign4speed}, the behavior of the slow-roll parameters and propagation speeds of scalar perturbations and gravitational waves in different approximations is shown. One can see that the approximation~II works better than the approximation~I at the end of inflation. In turn, even the approximation~I matches the numerical result much better than the standard approximation.


\begin{table}
\caption{\label{TablePhi4b} \textbf{Values of the inflationary parameters for the model with the quartic potential in different approximations.}}
\vspace{4mm}
\begin{tabular}{|c|c|c|c|}
\hline
\bf{Parameter} & \bf{Standard} & \bf{Approximation I} & \bf{Approximation II}\\
{} & \bf{approximation} & {} & {}\\
\hline
$\phi_{in}/M_\mathrm{Pl}$ & $2.5555$ & 1.4104 &  $1.4116$\\
 \hline
$10^9A_s(\phi_{in})$ & $2.10$ & $2.10$ & $2.10$\\
\hline
$n_s(\phi_{in})$ & $0.817$ & $0.964$ & $0.965$\\
\hline
$r(\phi_{in})$ & $0.0466$ & $0.0045$ & $0.0045$\\
\hline
$N(\phi_{end})-N(\phi_{in})$ & $13.5$ & $54.6$ &  $61.8$\\
\hline
\end{tabular}
\end{table}

\begin{figure}
\includegraphics[scale=0.41]{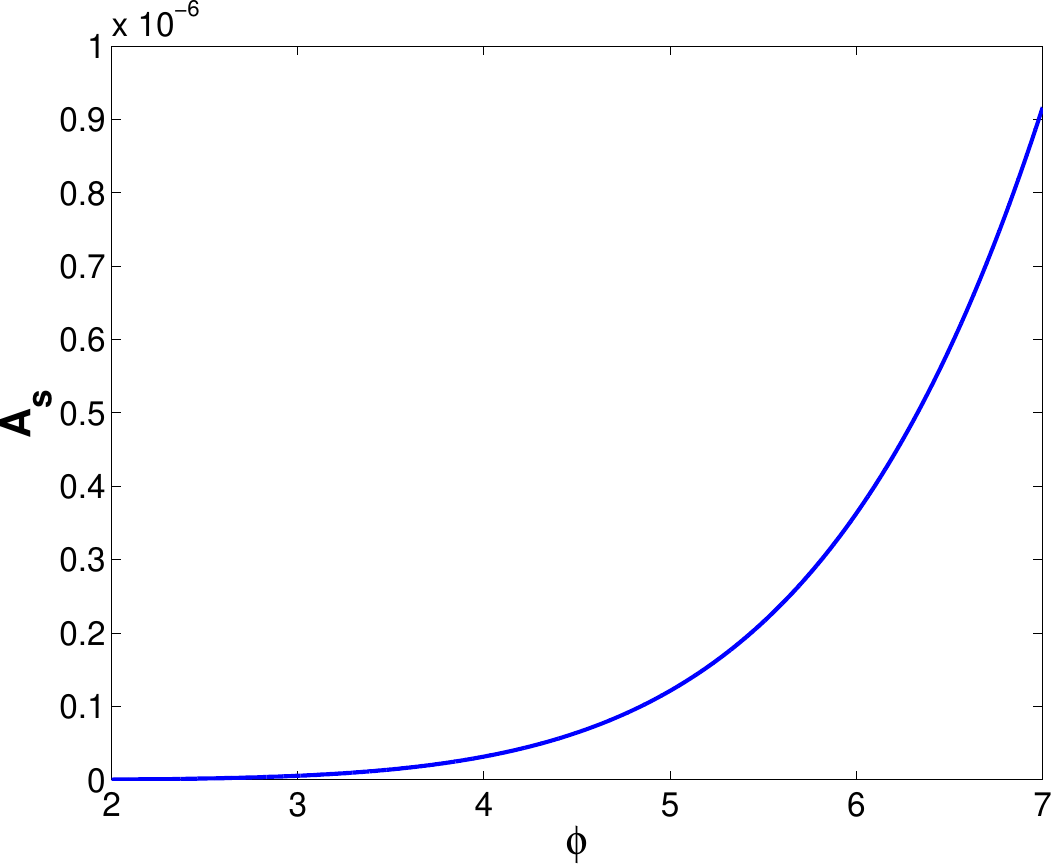}\quad
\includegraphics[scale=0.41]{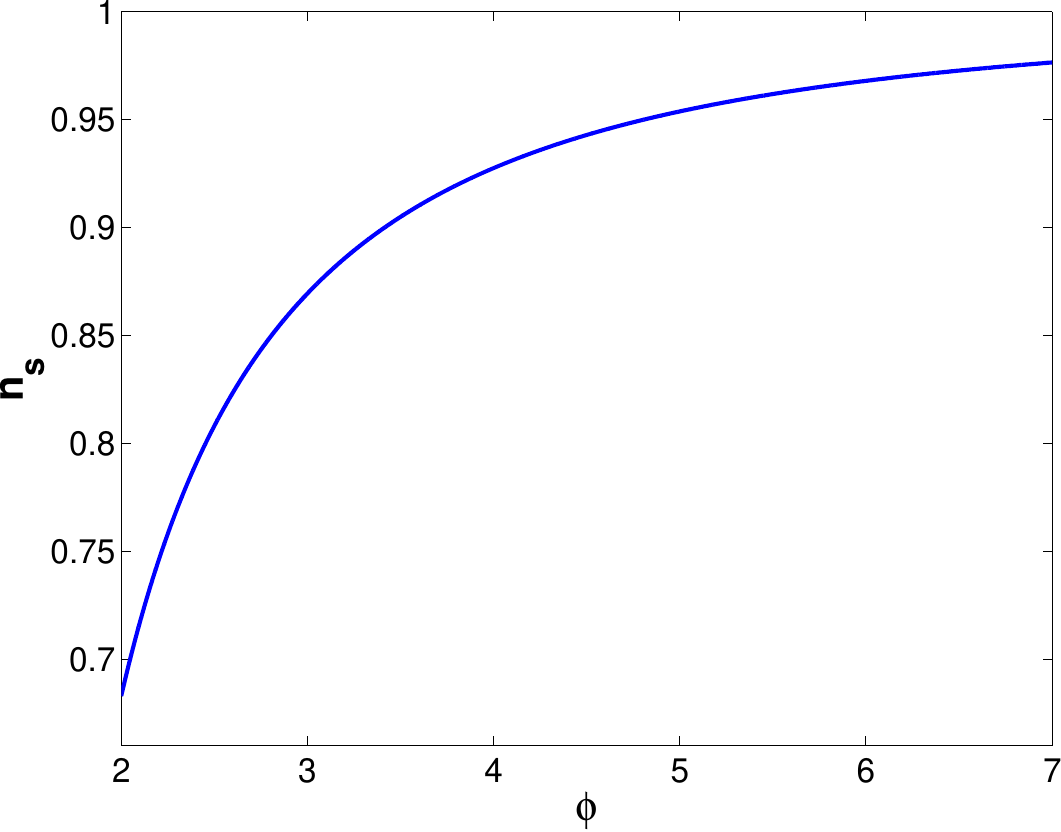}
\caption{The inflationary parameters $A_s(\phi)$ and $n_s(\phi)$ for the model with $V(\phi)=V_0\phi^4$ in the standard approximation. Values of $\phi$ are given in units of $M_\mathrm{Pl}$.}
\label{Fig3phi4}
\end{figure}

In the model with the quartic potential, we also check the possibility to get the suitable values of inflationary parameters using the same procedures as in the case of the quadratic potential. Solving the equation $A_s(\phi_{in})=2.1\times10^{-9}$, get $\phi_{in}$ and calculate values of $n_s(\phi_{in})$ and $r(\phi_{in})$ in all approximations. One can see in Table~\ref{TablePhi4b} that the standard approximation seriously underestimates the number of e-foldings. Looking at Fig.~\ref{Fig3phi4}, we see that in the standard approximation, either the value of $A_s(\phi_{in})$ is too large or the value of $n_s$ is too small. So, the standard approximation does not work, whereas both proposed approximations give acceptable values of the inflationary parameters (see Table~\ref{TablePhi4b}).

\section{Conclusion}
~~~~We consider different slow-roll approximations for inflationary models with the Gauss-Bonnet term and compare them with numerical solutions without any approximation.

The slow-roll parameters (\ref{epsilon}) and (\ref{delta}) depend on both the scalar field and the Hubble parameter. To construct a slow-roll approximation one should present the Hubble parameter as a function of the scalar field. The standard way~\cite{Guo:2010jr,Pozdeeva:2020apf,Odintsov:2023lbb} is to use Eq.~(\ref{Equ0lo}) that is the same relation for $H(\phi)$ as in inflationary GR models with a minimally coupled scalar field, whereas the slow-roll equation $\dot\phi=F(\phi)$ is different from the GR case. It has been noted on the example with the quartic potential and the inverse proportional function $\xi$ that the standard approximation is not correct at the end of inflation~\cite{vandeBruck:2015gjd}.

In this paper, we have shown on examples with quadratic and quartic potentials and the function $\xi$ given by formula~(\ref{xiphi})
that the standard approximation and numerical calculations give essentially different values of the scalar field during inflation. By this reason, the values of inflationary parameters calculated in the standard approximation are beyond the range of values admitted by observation, whereas numerical calculations show that the model does not contradict to observation data. To solve this discrepancy we propose two more accurate slow-roll approximations.

The construction of a higher accuracy slow-roll approximation is based on the use of not the function $H(\phi)$, given by Eq.~(\ref{Equ0lo}), but the function $H(\phi,\delta_1)$, given by Eq.~(\ref{equ00slr}) in the approximation~I or given by Eq.~(\ref{apprIIH2}) in the approximation~II. So, to get $H(\phi)$ we need to obtain $\delta_1(\phi)$. The knowledge of $\delta_1(\phi)$ also allows us to obtain $\chi(\phi)$ and, therefore, the function $\phi(N)$ as a solution of the first-order autonomic differential equation. On the next step, we get $\varepsilon_1(\phi)$, using $\frac{dH}{dN}=\chi H_{,\phi}$, where the values of $H(\phi)$ and  $\chi(\phi)$ are obtained in the corresponding approximation.

We have compared the proposed slow-roll approximations with the standard approximation and numerical calculations on models with power-law potentials $V=V_0\phi^n$, in the cases of $n=2$ and $n=4$.
Calculating the corresponding slow-roll parameters by direct numerical integration of background equations of motion without any approximation, we have found such values of parameters of the model that the amplitude of scalar perturbations and spectral index reproduce observed values and the tensor-to-scalar ratio is below the observation bound. For these values of the model parameters, the standard approximation is not accurate enough to get correct values of inflationary parameters and correct number of e-folding during inflation. On the contrary, the proposed more involved approximations give the results close enough to the numerical solutions. Observational parameters calculated using these approximations are still within the allowed regions for both model considered here.

In both our examples, the approximation~II works better than the approximation~I. Both approximations lose their validity in the limit $\delta_1\rightarrow 1$. It would be interesting to compare new approximations on models with $\delta_1$ close to unity during the slow-roll inflation. In principle, in certain cases, the approximation~I could work better than the approximation~II at $\delta_1\rightarrow 1$, because the value of $H^2$ does not tend to infinity, but it requires further investigations.

To the best of our knowledge, models with power-law potentials and the coupling function $\xi(\phi)$ given by Eq.~(\ref{xiphi}) have not been considered before. As a byproduct, we propose new inflationary models that are in agreement with the observation data and, thus, confirm earlier results that single scalar field inflationary models with quadratic and quartic potentials can match with observable data if  appropriate couplings to the Gauss-Bonnet term are added.

We plan to use the proposed approximations to analyze the known inflationary models and construct new models with the Gauss-Bonnet term.

\acknowledgments
A.T. is supported by
Russian Government Program of Competitive Growth of Kazan Federal University.

\appendix
\section{Explicit formulae for $\varepsilon_2(\phi)$ and $\delta_2(\phi)$}
\label{App}
\subsection{Standard Approximation}
~~~~From Eq.~(\ref{slrVeffe}) and \eqref{eps2delta2phi}, we obtain
\begin{equation}
\label{slrVeff2}
\begin{split}
\varepsilon_2&={}-\frac{2V}{U_0}{V_{eff}}_{,\phi}\frac{d}{d\phi}\ln\left(V_{,\phi}{V_{eff}}_{,\phi}\right)=
{}-\frac{2V}{U_0}{V_{eff}}_{,\phi}\left[\frac{V_{,\phi\phi}}{V_{,\phi}}+\frac{{V_{eff}}_{,\phi\phi}}{{V_{eff}}_{,\phi}}\right],\\
\delta_2&= { } -\frac{2V}{U_0}{V_{eff}}_{,\phi}\frac{d}{d\phi}\ln\left(V^2\xi_{,\phi}{V_{eff}}_{,\phi}\right).
\end{split}
\end{equation}

\subsection{Approximation I}
~~~~Using Eqs.~\eqref{eps2delta2phi} and \eqref{apprIeps1phi}, we get
\begin{equation}
\begin{array}{l}
\label{apprIeps2phi}
\varepsilon_2(\phi)=\frac{U_0\delta_1}{2\xi_{,\phi}H^2\varepsilon_1}{\varepsilon_1}_{,\phi}=\chi\left[\frac{d\ln(\chi)}{d\phi}+\frac{d}{d\phi}\left(\ln\left[\frac{d\ln\left(H^2\right)}{d\phi}\right]\right) \right]={}-\frac{6U_0^2\left(3U_0^2V_{,\phi}+\xi_{,\phi}V^2\right)}{V\left(9U_0^3-6U_0^2\xi_{,\phi}V_{,\phi}+\xi_{,\phi}^2V^2\right)}\\
\\\times\left[\frac{3{U_0}V_{,\phi\phi}+\xi_{,\phi\phi}V^2+2\xi_{,\phi}VV_{,\phi}}{3U_0^2V_{,\phi}+\xi_{,\phi}V^2}-\frac{V_{,\phi}}{V}+\frac{6U_0^2\xi_{,\phi\phi}V_{,\phi}+6U_0^2\xi_{,\phi}V_{,\phi\phi}
-2\xi_{,\phi}\xi_{,\phi\phi}V^2-2\xi_{,\phi}^2VV_{,\phi}}{9U_0^3-6U_0^2\xi_{,\phi}V_{,\phi}+\xi_{,\phi}^2V^2}+\right. \\
\\\left. +\left[{\frac{V_{,\phi}}{V}-\frac{6U_0^2\xi_{,\phi\phi}V_{,\phi}+6U_0^2\xi_{,\phi}V_{,\phi\phi}-2\xi_{,\phi}\xi_{,\phi\phi}V^2-2\xi_{,\phi}^2VV_{,\phi}}{9U_0^3-6U_0^2\xi_{,\phi}V_{,\phi}+\xi_{,\phi}^2V^2}-\frac{2\xi_{,\phi}\xi_{,\phi\phi}V^2    +2\xi_{,\phi}^2VV_{,\phi}}{3U_0^3+\xi_{,\phi}^2V^2}}\right]^{-1}\right.\\
\\\left.\times\left(\frac{V_{,\phi\phi}}{V}-\frac{V_{,\phi}^2}{V^2}-\frac{{\left(-6U_0^2\xi_{,\phi\phi}V_{,\phi}-6U_0^2\xi_{,\phi}V_{,\phi\phi}+2\xi_{,\phi}\xi_{,\phi\phi}V^2+2\xi_{,\phi}^2VV_{,\phi}\right)}^2}{{\left(9U_0^3-6U_0^2\xi_{,\phi}V_{,\phi}+\xi_{,\phi}^2V^2\right)}^2}\right.\right.\\
\\\left.\left. +\frac{-6U_0^2\xi_{,\phi\phi\phi}V_{,\phi}-12U_0^2\xi_{,\phi\phi}V_{,\phi\phi}-6U_0^2\xi_{,\phi}V_{,\phi\phi\phi}+2\xi_{,\phi\phi}^2V^2+2\xi_{,\phi}\xi_{,\phi\phi\phi}V^2+8\xi_{,\phi}\xi_{,\phi\phi}VV_{,\phi}+2\xi_{,\phi}^2V_{,\phi}^2+2\xi_{,\phi}^2VV_{,\phi\phi}}{9U_0^3-6U_0^2\xi_{,\phi}V_{,\phi}+\xi_{,\phi}^2V^2}
\right.\right. \\
\\\left.\left. {} -\frac{2\xi_{,\phi\phi}^2V^2+2\xi_{,\phi}\xi_{,\phi\phi\phi}V^2+8\xi_{,\phi}\xi_{,\phi\phi}VV_{,\phi}+2\xi_{,\phi}^2V_{,\phi}^2+2\xi_{,\phi}^2VV_{,\phi\phi}}{3U_0^3+\xi_{,\phi}^2V^2}
+\frac{{\left(2\xi_{,\phi}\xi_{,\phi\phi}V^2+2\xi_{,\phi}^2VV_{,\phi}\right)}^2}{\left(3U_0^3+\xi_{,\phi}^2V^2\right)^2}\right)\right].
\end{array}
\end{equation}

    Using Eqs.~\eqref{eps2delta2phi} and \eqref{apprIIdel1phi}, we obtain
\begin{equation}
\label{apprIdelta2phi}
\begin{split}
\delta_2(\phi)&={}-\frac{6U_0^2(3U_0^2V_{,\phi}+\xi_{,\phi}V^2)}{V(9U_0^3-6U_0^2\xi_{,\phi}V_{,\phi}+\xi_{,\phi}^2V^2)}\\
 &\times\left(\frac{\xi_{,\phi\phi}}{\xi_{,\phi}}+\frac{3U_0^2V_{,\phi\phi}+\xi_{,\phi\phi}V^2+2\xi_{,\phi}VV_{,\phi}}{3U_0^2V_{,\phi}+\xi_{,\phi}V^2}-\frac{2\xi_{,\phi}\xi_{,\phi\phi}V^2+2\xi_{,\phi}^2VV_{,\phi}}{3U_0^3+\xi_{,\phi}^2V^2}\right).
\end{split}
\end{equation}

\subsection{Approximation II}
~~~~From Eqs.~\eqref{eps2delta2phi} and \eqref{apprIIeps1phi}, we get
\begin{equation}
\begin{array}{l}
\label{apprIIeps2phi}
\varepsilon_2(\phi)=\frac{U_0\delta_1}{2\xi_{,\phi}H^2\varepsilon_1}{\varepsilon_1}_{,\phi}={}-\frac{2\left(3U_0^2V_{,\phi}+\xi_{,\phi}V^2\right)\left(9U_0^3-3U_0^2\xi_{,\phi}V_{,\phi}+2\xi_{,\phi}^2V^2\right)}{27U_0^2V{(U_0-\xi_{,\phi}V_{,\phi})}^2}\\
\\\times\left[\frac{3U_0^2V_{,\phi\phi}+\xi_{,\phi\phi}V^2+2\xi_{,\phi}VV_{,\phi}}{3U_0^2V_{,\phi}+\xi_{,\phi}V^2}+\frac{-3U_0^2\xi_{,\phi\phi}V_{,\phi}-3U_0^2\xi_{,\phi}V_{,\phi\phi}+4\xi_{,\phi}\xi_{,\phi\phi}V^2+4\xi_{,\phi}^2VV_{,\phi}}{9U_0^3-3U_0^2\xi_{,\phi}V_{,\phi}+2\xi_{,\phi}^2V^2}
-\frac{V_{,\phi}}{V}\right.\\
\\{}+\frac{2(\xi_{,\phi\phi}V_{,\phi}+\xi_{,\phi}V_{,\phi\phi})}{U_0-\xi_{,\phi}V_{,\phi}} +\left[\frac{V_{,\phi}}{V}-\frac{\xi_{,\phi\phi}V_{,\phi}+\xi_{,\phi}V_{,\phi\phi}}{U_0-\xi_{,\phi}V_{,\phi}} -\frac{4\xi_{,\phi}\xi_{,\phi\phi}V^2+4\xi_{,\phi}^2VV_{,\phi}-3U_0^2\xi_{,\phi\phi}V_{,\phi}-3U_0^2\xi_{,\phi}V_{,\phi\phi}}{9U_0^3-3U_0^2\xi_{,\phi}V_{,\phi}+2\xi_{,\phi}^2V^2}\right]^{-1}
\\
\\\left.\times\left(\frac{V_{,\phi\phi}}{V}-\frac{V_{,\phi}^2}{V^2}-\frac{\xi_{,\phi\phi\phi}V_{,\phi}+2\xi_{,\phi\phi}V_{,\phi\phi}+\xi_{,\phi}V_{,\phi\phi\phi}}{U_0-\xi_{,\phi}V_{,\phi}}-\frac{{(\xi_{,\phi\phi}V_{,\phi}+\xi_{,\phi}V_{,\phi\phi})}^2}{{(U_0-\xi_{,\phi}V_{,\phi})}^2}\right.\right.\\
\\\left.\left.{} -\frac{-3U_0^2\xi_{,\phi\phi\phi}V_{,\phi}-6U_0^2\xi_{,\phi\phi}V_{,\phi\phi}-3U_0^2\xi_{,\phi}V_{,\phi\phi\phi}+4\xi_{,\phi\phi}^2V^2+4\xi_{,\phi}\xi_{,\phi\phi\phi}V^2+16\xi_{,\phi}\xi_{,\phi\phi}VV_{,\phi}+4\xi_{,\phi}^2V_{,\phi}^2+4\xi_{,\phi}^2VV_{,\phi\phi}}{9U_0^3-3U_0^2\xi_{,\phi}V_{,\phi}+2\xi_{,\phi}^2V^2}\right.\right.\\
\\\left.\left.{}+\frac{{\left(4\xi_{,\phi}\xi_{,\phi\phi}V^2+4\xi_{,\phi}^2VV_{,\phi}-3U_0^2\xi_{,\phi\phi}V_{,\phi}-3U_0^2\xi_{,\phi}V_{,\phi\phi}\right)}^2}{\left(9U_0^3-3U_0^2\xi_{,\phi}V_{,\phi}+2\xi_{,\phi}^2V^2\right)^2}\right)\right].
\end{array}
\end{equation}

Using Eq.~(\ref{apprIIdel1phi}), we get
\begin{equation}
\label{apprIIdelta2}
\begin{split}
\delta_2(\phi)&={}-\frac{2(3U_0^2V_{,\phi}+\xi_{,\phi}V^2)(9U_0^3-3U_0^2\xi_{,\phi}V_{,\phi}+2\xi_{,\phi}^2V^2)}{27U_0^2V{(U_0-\xi_{,\phi}V_{,\phi})}^2}\\
\\&\times\left(\frac{\xi_{,\phi\phi}}{\xi_{,\phi}}+\frac{3U_0^2V_{,\phi\phi}+\xi_{,\phi\phi}V^2+2\xi_{,\phi}VV_{,\phi}}{3U_0^2V_{,\phi}+\xi_{,\phi}V^2}-\frac{\xi_{,\phi\phi}V_{,\phi}+\xi_{,\phi}V_{,\phi\phi}}{\xi_{,\phi}V_{,\phi}-U_0}\right).
\end{split}
\end{equation}

\bibliographystyle{JHEP}
\bibliography{BibliographyGBInflation}{}

\end{document}